\documentclass[11pt,draftcls,onecolumn] {IEEEtran}%1029
\usepackage{cite}% *** CITATION PACKAGES ***
\usepackage[cmex10]{amsmath}% *** MATH PACKAGES ***
\usepackage{amssymb}
\usepackage{amsthm}
\usepackage{mathrsfs}%SF
\usepackage{bm}
\usepackage[mathscr]{eucal}
\usepackage{amssymb,amsmath,amsthm,amsfonts,latexsym}
\usepackage{amsmath,graphicx,bm,xcolor,url}
\usepackage{graphicx}
\usepackage{latexsym}%SF
\usepackage{CJK}%SF
\usepackage{indentfirst}%SF
\usepackage{geometry}%SF
\usepackage{psfrag}%SF
\usepackage{setspace}%SF
\usepackage{algorithmic}% *** SPECIALIZED LIST PACKAGES ***
\usepackage{algorithmic, cite}
\usepackage{algorithm}
\usepackage{array}% *** ALIGNMENT PACKAGES ***
\usepackage{mdwmath}
\usepackage{mdwtab}
\usepackage{eqparbox}
\usepackage{url}% *** PDF, URL AND HYPERLINK PACKAGES ***
\usepackage{epstopdf}
\usepackage{epsfig,epsf,psfrag}
\usepackage{fixltx2e}%ordering of single and double column floats
\usepackage{verbatim}
\usepackage{textcomp}
\usepackage{psfrag} %%25 June

\usepackage{caption}
\usepackage{subcaption}%2013-03-16

\theoremstyle{plain}
\newtheorem{mythe}{Theorem}
\theoremstyle{remark}
\newtheorem{mylem}{Lemma}
\theoremstyle{plain}

\theoremstyle{remark}

\theoremstyle{plain}

\theoremstyle{remark}
\newtheorem{myrem}{Remark}
\theoremstyle{remark}
\newtheorem{myexa}{Example}
\theoremstyle{remark}

\theoremstyle{remark}

\theoremstyle{remark}

\theoremstyle{remark}

\catcode`~=11 \def\UrlSpecials{\do\~{\kern -.15em\lower .7ex\hbox{~}\kern .04em}} \catcode`~=13

\allowdisplaybreaks[4]

\newcommand{\calC}{\mathcal{C}}
\newcommand{\calD}{\mathcal{D}}

\newcommand{\calI}{\mathcal{I}}

\newcommand{\calK}{\mathcal{K}}

\newcommand{\calN}{\mathcal{N}}

\newcommand{\calQ}{\mathcal{Q}}

\newcommand{\calS}{\mathcal{S}}

\newcommand{\bA}{\mathbf{A}}

\newcommand{\bB}{\mathbf{B}}

\newcommand{\bC}{\mathbf{C}}
\newcommand{\bd}{\mathbf{d}}
\newcommand{\bD}{\mathbf{D}}
\newcommand{\be}{\mathbf{e}}
\newcommand{\bE}{\mathbf{E}}

\newcommand{\bF}{\mathbf{F}}

\newcommand{\bh}{\mathbf{h}}

\newcommand{\bI}{\mathbf{I}}

\newcommand{\boldm}{\mathbf{m}}

\newcommand{\bR}{\mathbf{R}}
\newcommand{\bs}{\mathbf{s}}

\newcommand{\bu}{\mathbf{u}}
\newcommand{\bU}{\mathbf{U}}

\newcommand{\bw}{\mathbf{w}}

\newcommand{\bx}{\mathbf{x}}
\newcommand{\bX}{\mathbf{X}}
\newcommand{\by}{\mathbf{y}}

\newcommand{\bz}{\mathbf{z}}

\newcommand{\bbC}{\mathbb{C}}

\newcommand{\bbE}{\mathbb{E}}

\newcommand{\bbR}{\mathbb{R}}

\DeclareMathAlphabet{\mathbsf}{OT1}{cmss}{bx}{n}
\DeclareMathAlphabet{\mathssf}{OT1}{cmss}{m}{sl}% slanted sans serif

\DeclareSymbolFont{bsfletters}{OT1}{cmss}{bx}{n}
\DeclareSymbolFont{ssfletters}{OT1}{cmss}{m}{n}
\DeclareMathSymbol{\bsfGamma}{0}{bsfletters}{'000}
\DeclareMathSymbol{\ssfGamma}{0}{ssfletters}{'000}
\DeclareMathSymbol{\bsfDelta}{0}{bsfletters}{'001}
\DeclareMathSymbol{\ssfDelta}{0}{ssfletters}{'001}
\DeclareMathSymbol{\bsfTheta}{0}{bsfletters}{'002}
\DeclareMathSymbol{\ssfTheta}{0}{ssfletters}{'002}
\DeclareMathSymbol{\bsfLambda}{0}{bsfletters}{'003}
\DeclareMathSymbol{\ssfLambda}{0}{ssfletters}{'003}
\DeclareMathSymbol{\bsfXi}{0}{bsfletters}{'004}
\DeclareMathSymbol{\ssfXi}{0}{ssfletters}{'004}
\DeclareMathSymbol{\bsfPi}{0}{bsfletters}{'005}
\DeclareMathSymbol{\ssfPi}{0}{ssfletters}{'005}
\DeclareMathSymbol{\bsfSigma}{0}{bsfletters}{'006}
\DeclareMathSymbol{\ssfSigma}{0}{ssfletters}{'006}
\DeclareMathSymbol{\bsfUpsilon}{0}{bsfletters}{'007}
\DeclareMathSymbol{\ssfUpsilon}{0}{ssfletters}{'007}
\DeclareMathSymbol{\bsfPhi}{0}{bsfletters}{'010}
\DeclareMathSymbol{\ssfPhi}{0}{ssfletters}{'010}
\DeclareMathSymbol{\bsfPsi}{0}{bsfletters}{'011}
\DeclareMathSymbol{\ssfPsi}{0}{ssfletters}{'011}
\DeclareMathSymbol{\bsfOmega}{0}{bsfletters}{'012}
\DeclareMathSymbol{\ssfOmega}{0}{ssfletters}{'012}

\newcommand{\hatE}{\widehat{E}}

\newcommand{\tilE}{\widetilde{E}}

\newcommand{\hath}{\widehat{h}}

\newcommand{\hatbh}{\widehat{\bh}}

\newcommand{\tilv}{\widetilde{v}}
\newcommand{\tilV}{\widetilde{V}}

\newcommand{\tilbx}{\widetilde{\bx}}

\newcommand{\barE}{\bar{E}}

\newcommand{\barQ}{\bar{Q}}

\newcommand{\bGamma}{\bm{\Gamma}}

\newcommand{\bSigma	}{\bm{\Sigma}}

\def\norm#1{\left\| #1 \right\|}
\def\norm2#1{\left\| #1 \right\|_2}
\def\norm22#1{\left\| #1 \right\|_2^2}

\newcommand{\eqa}{\stackrel{(a)}{=}}
\newcommand{\eqb}{\stackrel{(b)}{=}}
\newcommand{\eqc}{\stackrel{(c)}{=}}

\DeclareMathOperator{\tr}{tr}

\newcommand{\qednew}{\nobreak \ifvmode \relax \else
      \ifdim\lastskip<1.5em \hskip-\lastskip
      \hskip1.5em plus0em minus0.5em \fi \nobreak
      \vrule height0.75em width0.5em depth0.25em\fi}

\allowdisplaybreaks[1]
\usepackage{float}
\usepackage{bbm}
\title{Dynamic Resource Allocation for Multiple-Antenna Wireless Power Transfer}
\author{Gang~Yang, Chin~Keong~Ho, and Yong~Liang~Guan  % <this % stops a space
\thanks{G.~Yang and Y.~L.~Guan are with the School of Electrical and Electronic Engineering, Nanyang Technological University, Singapore (e-mail:\{yang0305, eylguan\}@ntu.edu.sg).} % <this % stops a space
\thanks{C. K. Ho is with the Institute for Infocomm Research, A$^\star$STAR, Singapore (e-mail: hock@i2r.a-star.edu.sg). }}
\begin{document}
\maketitle %Automatic title!

\vspace{-0.5in}
\begin{abstract}
We consider a point-to-point multiple-input-single-output (MISO) system where a receiver harvests energy from a wireless power transmitter to power itself for various applications. To achieve high-efficiency wireless power transfer, the transmitter performs {\emph{energy beamforming}} by using an instantaneous channel state information (CSI). The CSI is estimated at the receiver by training via a preamble, and fed back to the transmitter. The channel estimate is more accurate when longer preamble is used, but less time is left for wireless power transfer before the channel changes. To maximize the harvested energy, in this paper, we address the key challenge of balancing the time resource used for channel estimation and wireless power transfer, and also investigate the allocation of energy resource used for wireless power transfer.
First, we consider the general scenario where the preamble length is allowed to vary dynamically. Taking into account the effects of imperfect CSI, the optimal preamble length is obtained online by solving a dynamic programming (DP) problem. The solution is shown to be a threshold-type policy that depends only on the channel estimate power (i.e., the squared $l_2$-norm of the channel estimate). Next, we consider the scenario in which the preamble length is fixed. The optimal preamble length is optimized offline. Furthermore, we derive the optimal power allocation schemes for both scenarios. For the scenario of dynamic-length preamble, the power is allocated according to both the optimal preamble length and the channel estimate power; while for the scenario of fixed-length preamble, the power is allocated according to only the channel estimate power. The analysis results are validated by numerical simulations. Encouragingly, with optimal power allocation, the harvested energy by using optimized fixed-length preamble is almost the same as the harvested energy by employing dynamic-length preamble, hence allowing a low-complexity wireless power transfer system to be implemented in practice.
%we derive the optimal energy beamformer for Rayleigh channels,
%sufficient to approach
% and further maximize the net information rate achieved by using the harvested Finally, we maximize the net information rate achieved by using the harvested energy. %For this problem, we consider two typical application scenarios, wherein we maximize the net average information rate, or maximize the sensing accuracy.
%both correlated and uncorrelated
%For the scenario of dynamic-length preamble,
%Using a least square (LS) channel estimator in uncorrelated channels, we obtain the optimal length of training preamble to maximize the expected harvested wireless power.
\end{abstract}

\newpage

\section{Introduction}
Recently, far-field wireless power transfer (WPT) has emerged as a promising technology to address energy and lifetime bottlenecks for power-limited devices in wireless networks~\cite{Varshney08, GroverShannonTesla10, SharmaEnergyMangEHC10}. For example, in an energy harvesting sensor network, sensors can harvest energy to power themselves. The harvested energy is used for data transmission by various schemes, such as wireless compressive sensing proposed in~\cite{YangTanHo13}. Since electromagnetic (EM) waves decay quickly over distance, the EM waves have to be concentrated into a narrow beam to achieve efficient power transfer. This is referred to as {\textit{energy beamforming}}~\cite{MIMOWIPTZhang13}, which was first considered for simultaneous wireless information and power transfer (SWIPT) in multiuser downlink~\cite{MIMOWIPTZhang13}. Assuming perfect channel state information (CSI) at the transmitter, \cite{WITOppoEHLiuZhangChua13} investigated joint optimization of transmit power control, information and power transfer scheduling; \cite{KHuangELarsson13} studied resource allocation algorithms for SWIPT in broadband wireless systems.
%; \cite{HJuRZhang13} considered a multi-antenna SWIPT system without CSIT.
%, where the transmitter uses random beamforming technique to generate artificial channel fading to enable efficient energy harvesting
%In~\cite{ANasirAKennedy12}, the throughput was analyzed for both TS-based amplify-and-forward (AF) relaying systems and PS-based AF relaying systems, under the assumption of perfect CSI at the receiver.
%, where the time-switching (TS) and power-splitting (PS) receiver schemes are proposed
%WPT refers to using the radiative electromagnetic (EM) wave emitted from a power source to deliver energy to a receiver. The energy receiver typically uses a rectifying-antenna circuit, or rectenna in short, to harvest radio-frequency (RF) signals directly~\cite{BStrassnerKZhang13}.

%Since the energy of radiative electromagnetic (EM) waves decay quickly over distances, the EM energy can be concentrated to achieve efficient transmission of power, also referred to as energy beamforming~\cite{MIMOWIPTZhang13}.

%Since the energy of radiative electromagnetic (EM) waves decay quickly over distances, the EM energy can be concentrated to achieve efficient transmission of power, also referred to as energy beamforming~\cite{MIMOWIPTZhang13}.

%On the other hand,
The uplink wireless information transfer (WIT) powered by downlink WPT  was considered in~\cite{HJuRZhang13,XChenCYuen13}. A harvest-then-transmit protocol was proposed in~\cite{HJuRZhang13}, where all users first harvest the wireless energy in the DL and then send independent information in the uplink by time division multiple access. With perfect CSI, the sum throughput was maximized by jointly optimizing the time allocation for the downlink WPT and uplink WIT. ~\cite{XChenCYuen13} considered the single-user scenario, where the optimal time duration for downlink WPT is determined to maximize an approximate lower bound of the uplink information rate.
%\cite{KHuangVLau13} optimized the transmit power and deployment density of a cellular network overlaid with wireless powering mobiles, to guarantee an outage probability of information and power transfer. %\cite{SLeeKHuang13} maximized the secondary network throughput in a cognitive radio network powered by opportunistic wireless energy harvesting. %an upper bound and

%\cite{XChenCYuen13} shows that an upper bound on the average information rate decreases as the CSI accuracy decreases.
The knowledge of CSI is an essential prerequisite for both energy beamforming and information decoding. For instance, \cite{WITOppoEHLiuZhangChua13} showed that the rate-energy tradeoff in SWIPT systems degrades as the CSI accuracy decreases. Typically, the receiver needs to perform channel estimation and feed back CSI to the transmitter before power transfer. In practice, perfect CSI at the transmitter is not available due to various factors such as time-varying channel, inaccurate channel estimation, quantization error and feedback error. When the channel uncertainty is considered as deterministic and norm bounded, robust beamforming design was studied in~\cite{ZXiangMTao12} for a multiple-input-single-output (MISO) system with SWIPT, in~\cite{DLiZQiu13} for a two-way relay system with SWIPT. In~\cite{ZXiangMTao12}, the harvested energy was maximized for the worst channel realization, while guaranteeing that the information rate is above a threshold for all possible channel realizations. However, the actual worst case may occur with a very low probability. Hence, this worst-case approach may be overly conservative and therefore, lead to unnecessary performance degradation. %There are thus some works on WPT with imperfect CSI at the transmitter.
%\cite{DLiZQiu13} maximized the weighted sum rate under the power constraints at relay nodes and the energy harvesting constraints at the two source nodes.

Energy beamforming based on more accurate CSI contributes to higher efficiency of power transfer. The receiver, however incurs significant time (overhead) to obtain the accurate CSI. Longer time duration for channel estimation denotes more accurate CSI available at the transmitter, but also shortens the WPT duration, which may lead to less harvested energy. To maximize the harvested energy, there is thus a design freedom, namely the time spent for estimating the channel. Moreover, to improve the overall system energy efficiency, the amount of energy used for WPT should be optimized, for example, less energy is used for severely-fading channels. However, we are not aware of any work that takes into account the preamble overhead and energy allocation for wireless power transfer in a wireless (communication) system.

We consider a frame-based MISO system in which the transmitter performs energy beamforming using imperfect CSI fed back from the receiver. In this paper, we focus on efficient wireless power transfer; and put particular use of the harvested energy such as uplink wireless information transmission as future work extension \footnote{The receiver can use the harvested energy to perform various applications such as wireless transmission and sensing. For the application of uplink wireless transmission using the harvested energy, interested readers may refer to ~\cite{HJuRZhang13,XChenCYuen13}.}, to avoid obscuring this work. The frame is divided into four phases as shown in Fig.~\ref{fig:Fig0}: the channel estimation (CE) phase, the feedback phase, the wireless power transfer (WPT) phase, as well as the general energy utilization (EU) phase. The feedback is assumed to be error-free and take negligible time, and is thus ignored in the analysis. The time duration for the EU phase is fixed throughout. Unlike previous work on robust beamforming in~\cite{ZXiangMTao12, DLiZQiu13}, we maximize the harvested energy by balancing the time durations between the CE phase and the WPT phase, as well as allocating transmit power for the WPT phase.

To maximize the harvested energy, we consider two scenarios, namely dynamic-length preamble and fixed-length preamble, respectively. Given a channel estimate, we first derive the optimal energy beamformer, which applies to both scenarios. Then, we adjust the time duration for the CE phase. For the first scenario where the preamble length is allowed to vary dynamically, we obtain the optimal online preamble length by solving a dynamic programming (DP) problem. The solution is shown to be a threshold-type policy, wherein if the channel estimate power (i.e., the squared $l_2$-norm of the channel estimate) is less than a time-dependent threshold, the receiver continues to perform CE, and requests for wireless power otherwise. For the second scenario in which the preamble length is fixed for all frames, we optimize the preamble length offline. Moreover, we adjust the power allocated for WPT in each frame, for both scenarios. For the scenario of dynamic-length preamble, the power for WPT is allocated according to both the optimal preamble length and the channel estimate power; while for the scenario of fixed-length preamble, the power for WPT is allocated according to only the channel estimate power. Numerical results are finally given to validate our analysis.
%. Moreover, we derive the optimal power allocation schemes  and the WPT phase

The paper is organized as follows. In Section~\ref{SystemModel}, we describe the system model, and give the problem formulations. We study the optimal energy beamformer in Section~\ref{OptBeamforming}. In Section~\ref{WPT_MISO_Dynamic}, we allow the preamble length to vary with frames, and use dynamic programming to find the optimal preamble length. In Section~\ref{WPT_MISO_Static}, we fix the preamble length for all frames, and derive the optimal preamble length offline. Section~\ref{SecOPA} derives the optimal power allocation schemes. Section~\ref{Section_NumericalResults} gives numerical results.

\section{System Model} \label{SystemModel}
We consider a frame-based wireless power transfer system, consisting of a wireless power (WP) transmitter with $m$ antennas, a single-antenna receiver that is also known as a WP receiver, a downlink channel for wireless power transfer from the WP transmitter to the WP receiver, as well as a feedback channel to send CSI (and data) from the WP receiver to the WP transmitter. Hence, the WP transmitter and WP receiver also serve as the roles of the information receiver and information transmitter, respectively.
%in the case of feedback or data transmission from the WP receiver to the WP transmitter.

\begin{figure}[htbp]
\centering
\includegraphics[width=.65\columnwidth] {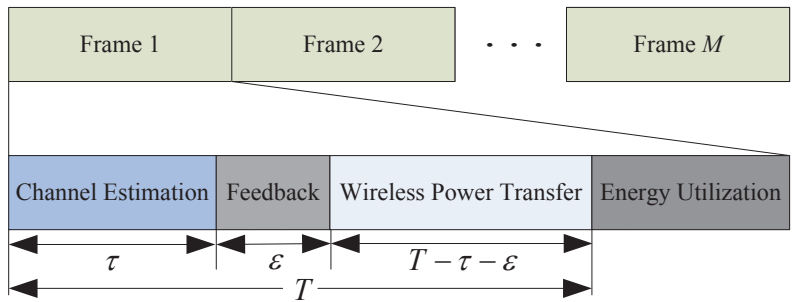}
\caption{Frame Structure}
\label{fig:Fig0}
\end{figure}

As shown in Fig.~\ref{fig:Fig0}, each frame consists of four phases. To focus on efficient wireless power transfer, the time duration for the fourth EU phase\footnote{In the EU phase, the harvested energy is used to by the receiver to perform various applications such as wireless transmissions.} is fixed in this paper. We assume the time duration for the CE, feedback and WPT phases in one frame is fixed as $T$ symbol periods. In the first CE phase, the WP transmitter sends preambles, and the WP receiver performs channel estimation in an interval of $\tau$ symbol periods. Then in the second phase, the WP receiver feeds the CSI back to the WP transmitter within $\epsilon$ symbol periods. In the third WPT phase, the WP transmitter delivers power via beamforming. The WP receiver harvests energy from the radio-frequency (RF) signals\footnote{For simplicity, we do not take interference into account; nevertheless our analysis can include random interference if statistical information of the interference is available.}. %For analytical convenience, we ignore the feedback time $\epsilon$, i.e., $\epsilon=0$.

We assume there is a lossless link for CSI feedback. For analytical simplicity, the feedback time is further assumed to be negligible and thus ignored, i.e., $\epsilon=0$. The downlink MISO channel $\bh$ is assumed to be quasi-static flat Rayleigh fading in each frame, i.e., $\bh \sim \calC \calN \left(\mathbf{0}_m, \bR\right)$, where $\mathbf{0}_m$ is the all-zero column vector of length $m$, and $\bR \triangleq \bbE \left(\bh \bh^{H} \right)$ denotes the $m \times m$ channel covariance matrix. We assume that $\bR$ is a full-rank matrix. The channel is referred to be uncorrelated if $ \bR =\bI_m$, where $\bI_m$ denotes the identity matrix of size $m$. The channel may vary independently from frame to frame.
%coefficients remain constant during each frame, but
%Therefore, the correlated MISO channel can be represented as
%\begin{align}
%  \bh = \bR^{1/2} \bh_w, \label{SpatialCorrelation}
%\end{align}
%where $\bh_w \sim \calC \calN \left(\mathbf{0}_m, \bI_m\right)$.

\subsection{Wireless Energy Beamforming}
We assume the time duration for CE and WPT can be divided into $N$ time slots, each of which consists of $m$ successive symbol periods, i.e., $T=mN$. The preamble that consists of $k=\frac{\tau}{m}$ time slots is used to obtain the channel estimate, denoted as $\hatbh_k$.

In the WPT phase, the received baseband signal in the $n$-th symbol period is written as
\begin{align}
  y_n = \tilbx_n^H \bh + z_n,\label{SignlModelMatWPT}
\end{align}
where $\tilbx_n$ is the $m \times 1$ transmitted signal vector, and $z_n \sim \calC \calN (0, \sigma_z^2)$ is the additive white Gaussian noise. Given channel estimate $\hatbh_{k}$, we denote the $m \times 1$ beamforming vector as $\bw (\hatbh_{k})$. The transmitted signal $\tilbx_n$ is then obtained as $\tilbx_n=\bw(\hatbh_{k})  \odot \bs_n$, where $\odot$ is the element-wise product, $\bs_n \in \bbC^{m}$ is a zero-mean and unit-power independent signal, i.e., $\bbE \left(\bs_n \bs_n^{H}\right) = \bI_m$. The subscript $n$ is ignored in the sequel.

Due to the law of energy conservation with efficiency $\rho$, the harvested RF-band energy normalized by the baseband symbol period, denoted by $E_0$, at the WP receiver is assumed to be proportional to that of the received baseband signal, i.e.,
\begin{align}
  E_0 = \rho \bbE_{\bh, \tilbx} \left( \left| \tilbx^H \bh \right|^2 \right) = \rho \bbE_{\bh, \hatbh_k} \left( \left|\bw^H(\hatbh_{k}) \bh \right|^2 \right),\label{EHEnergy1}
\end{align}
We assumed in \eqref{EHEnergy1} that the energy due to the ambient noise can not be harvested. For convenience, we also assume $\rho = 1$ in this paper. %and thus can be ignored Specifically, $\rho$ accounts for the loss in the energy transducer for converting the harvested energy to electrical energy to be stored; for convenience,

\subsection{Problem Formulation} \label{ProblemFormulation_MISO}
The WP receiver aims to harvest energy as much as possible in the WPT phase. Intuitively, longer preamble can increase the accuracy of channel estimation, and thus increase the efficiency of power transfer, but at the cost of reduced time left for the WPT phase. We also note that the power of the received signal depends on the fading condition in one frame. Hence, to maximize the harvested energy, we first obtain the optimal preamble length dynamically and offline, respectively, with constant transmit power for WPT; then we optimize the transmit power for WPT in each frame, namely the power allocation. Specifically, we consider wireless power transfer in two scenarios: dynamic-length preamble and fixed-length preamble, respectively.

\subsubsection{WPT with dynamic-length preamble}
We consider the scenario where the preamble length is allowed to vary dynamically, i.e., the receiver can decide to perform CE or request for WP at any time slot based on its current channel estimate. We denote the beginning of the $(k+1)$-th time slot as time instant $k$, where $k=0,1,\cdots,N-1$. At time instant $k=0$, the receiver decides to perform CE or request for WP in the first slot. If it decides to request for WP at $k=0$, the transmitter performs WPT in the first slot without beamforming. Otherwise, the transmitter sends preambles in the first slot, and the receiver obtains the channel estimate $\hatbh_1$ at the end of the first slot. For the subsequent time instant $k$, if the receiver decides to request for WP, it feeds back the channel estimate $\hatbh_k$ to the transmitter. Then the transmitter performs WPT using optimal beamformer $\bw^{\star}(\hatbh_k)$ in the next slot. If the receiver decides to continue CE at instant $k$, the transmitter sends preambles in the $(k+1)$-th time slot. The optimal beamformer $\bw^{\star}(\hatbh_k)$ will be found in Section~\ref{OptBeamforming}.

In Section~\ref{WPT_MISO_Dynamic}, we first formulate a dynamic programming (DP) problem to maximize the harvested energy in case that constant transmit power is used for WPT. We define therein the control space $\calC$, decision variable $u_k$ and the system state $\bx_k$. We define a policy as a sequence of functions $\mu_k(\bx_k)$ which maps each system state into a decision at time instant $k=0,1,\cdots,N-1$. The set of all possible policies is denoted as $\mathit{\Pi}$. Let $g_k(\bx_k,u_k)$ be the energy harvested in slot $k$ with state $\bx_k$ and decision $u_k$.
To maximize the expected harvested energy in all slots, we thus have the following optimization problem
\begin{align}
    \mathrm{(P_1)} \ \ \underset{\pi \in \mathit{\Pi}}{\text{max}} \ \   \bbE \left(\sum \nolimits_{k=0}^{N-1} g_k (\bx_k,u_k) \right) \label{eq:optimDP}
\end{align}
The expectation is performed over all random variables, specifically the channel $\bh$ and the channel estimates $\{\hatbh_k\}$ which become available only after the decision of CE or WP is made. The optimal policy $\pi^{\star}$ for $\mathrm{P_1}$ is obtained in Section~\ref{OptPolicy}.

Then, we derive the optimal power allocation scheme in Section~\ref{Sec2DPA}, which allocates transmit power for WPT according to both optimal preamble length adapted by employing the optimal policy $\pi^{\star}$, and the channel estimate power.%{DLPA}
%Also in Section~\ref{SecLAPA}, we obtain a low-complexity optimal power allocation scheme in which the transmit power for WPT is allocated according to only the optimal preamble length.

\subsubsection{WPT with fixed-length preamble}
To reduce implementation complexity, we consider the scenario in which the preamble length is fixed as $k$ time slots in all frames, but can be optimized offline. Then, the WPT phase in each frame consists of $(T-\tau)$ symbol periods, where $\tau=km$.

In Section~\ref{WPT_MISO_Static}, we first maximize the harvested energy in case that constant transmit power is used for WPT, by optimizing both the preamble length $k$ and the beamforming vector $\bw(\hatbh_{k})$. Specifically, we have the following optimization problem
\begin{subequations}
\label{eq:optimP1}
\begin{align}
%   \mathrm{(P_2)} \  \underset{\substack{\bw(\hatbh_{k}) \\  0 \leq k \leq N-1}} {\text{max}} \ \
   \mathrm{(P_2)} \  \underset{{\bw(\hatbh_{k}), \; 0 \leq k \leq N-1}} {\text{max}} \ \
%   \mathrm{(P_2)} \  \underset{\begin{subarray}{l}\bw(\hatbh_{k}) \\  0 \leq k \leq N-1 \end{subarray}} {\text{max}} \ \
   & %E(\tau, \bw(\hatbh_{\tau}) =
   (T-km)  \bbE_{\bh, \hatbh_{k}} \left( \left|\bw^{H}(\hatbh_{k}) \bh \right|^2 \right)  \label{eq:rewardP1} \\
    \qquad \text{s. t.} \ \
        & \left\| \bw(\hatbh_{k}) \right\|_2 = 1, \ \forall \ \hatbh_{k} \in \calC^m.  \label{eq:const1P1} %\\
        %& \tau=k m, \quad {\rm{for}} \ \ k \in \{1,2, \cdots, N\}. \  \label{eq:const2P1}
\end{align}
\end{subequations}
We will find the optimal solution $\bw^{\star}$ (as a function of $\hatbh_k$) and $k^{\star}$ in Section~\ref{OptBeamforming}, and Section~\ref{WPT_MISO_Static}, respectively.

Then, we derive the optimal power allocation scheme in Section~\ref{PA_FixedLength_LS}, which allocates transmit power for WPT according to only the channel estimate power.

\section{Optimal Energy Beamforming} \label{OptBeamforming}
In this section, we obtain the optimal beamforming vector $\bw_{\mathrm{opt}}(\hatbh)$, which shall be used to find the solutions to problem $(\mathrm{P_1})$, $(\mathrm{P_2})$ in Section~\ref{WPT_MISO_Dynamic}, Section~\ref{WPT_MISO_Static}, respectively.
%and $(\mathrm{P_3})$  and Section~\ref{Rate_Maximization}

\subsection{Partial or Full Feedback}
%The energy beamforming is performed at the transmitter, with imperfect CSI which is fed back from the receiver.
In practice, it is difficult for the transmitter to obtain full CSI due to the limited feedback capacity. This motivates us to investigate the impact of different amount of feedback on energy beamforming and thus the harvested energy. Unlike conventional limited feedback of quantized CSI, we let the receiver selectively feed back only $q$ largest, $1 \leq q \leq m$, {\textit{unquantized}} channel coefficients to the transmitter, so as to reduce the feedback amount. If $q=m$, it reduces to the conventional full CSI feedback. Let $\hath_{(i)}$ denote the channel coefficient with the $i$-th largest channel gain. The receiver feeds back the vector $\hatbh_q \triangleq \left[ \hath_{(1)}, \hath_{(2)}, \cdots, \hath_{(q)}\right]^T$ and the corresponding index set $\calI \triangleq \left\{(1), (2), \cdots, (q) \right\}$, to the transmitter. The parameter $q$ is defined as a metric, namely the {\textit{feedback dimension}}, to quantify the cost/amount of feedback.
The transmitter uses antennas with index in $\calI$ to perform energy beamforming.

\subsection{Optimal Energy Beamforming}
The energy beamforming is performed by using imperfect CSI at the transmitter. We first derive the distribution of the channel $\bh$ conditioned on a general {\textit{unbiased}} channel estimate $\hatbh$. We consider the $q$-dimensional feedback of CSI. Define the estimation error $\be_q \triangleq \hatbh_q - \bh_q$. Let $\bR_q \triangleq \bbE \left(\hatbh_q \hatbh_q^H \right) $ and $\bR_{\be,q}\triangleq \bbE \left(\be_q \be_q^H \right)$ be the $q$-dimensional counterparts of channel covariance matrix $\bR$ and the error covariance matrix $\bR_{\be}$, respectively. From equation (16) in Section IV-A in~\cite{IsertePalomar09}, we further have the following lemma.
\begin{mylem}\label{mylem1}
Let $\hatbh_q =\bh_q + \be_q$. Assume the channel vector $\bh_q \sim \calC \calN (\mathbf{0}_q, \bR_q)$, the error vector $\be_q \sim \calC \calN (\mathbf{0}_q, \bR_{\be,q})$, and $\bh_q$ and $\be_q$ are jointly Gaussian distributed. Given $\hatbh_q$, the vector $\bh_q$  follows a complex Gaussian distribution, i.e.,%joint to reflect partial or full feedback of the
\begin{align}
    \bh_q \left| \hatbh_q \right. \sim \calC \calN \left( \boldm_{\bh_q \left| \hatbh_q \right.},  \bSigma_{\bh_q \left| \hatbh_q\right.}\right)
\end{align}
where $ \boldm_{\bh_q \left| \hatbh_q \right.} =\left( \bR_{\be,q} \bR_q^{-1} + \bI_q \right)^{-1} \hatbh_q$, and $\bSigma_{\bh_q \left| \hatbh_q \right.} =\left( \bR_q^{-1} + \bR_{\be,q}^{-1} \right)^{-1}$.
\end{mylem}

%\begin{IEEEproof}
%See proofs in Appendix~\ref{AppCondPdf}.
%\end{IEEEproof}

From Lemma~\ref{mylem1}, the conditional correlation matrix is
\begin{align}
\bR_{\bh_q \left| \hatbh_q \right.} &= \bSigma_{\bh_q \left| \hatbh_q\right.} + \boldm_{\bh_q \left| \hatbh_q \right.} \boldm_{\bh_q \left| \hatbh_q \right.}^H. \label{CondCorreMtx}
\end{align}
Denote the singular value decomposition (SVD) of $\bR_{\bh_q \left| \hatbh_q \right.}$ by $\bR_{\bh_q \left| \hatbh_q \right.}=\bU_q \bGamma_q \bU_q^H$, where $\bU_q=[\bu_{1,q} \ \bu_{2,q} $ $ \cdots \ \bu_{q,q}]$, $\bGamma_q=\mathrm{diag} \{\gamma_1 \ \gamma_2 \ \cdots \ \gamma_q \}$, and $\gamma_1 \geq \gamma_2 \geq \cdots \geq \gamma_q$. We further have Lemma~\ref{LemmaOptBeamformer}.
%We further obtain the optimal beamformer in Lemma~\ref{LemmaOptBeamformer}.
\begin{mylem}\label{LemmaOptBeamformer}
Assume the channel vector $\bh_q \sim \calC \calN (\mathbf{0}_q, \bR_q)$, the error vector $\be_q \sim \calC \calN (\mathbf{0}_q, \bR_{\be,q})$, and $\bh_q$ and $\be_q$ are jointly Gaussian. Given $\hatbh_q$, the optimal beamforming vector that maximizes the normalized harvested energy, is given by
\begin{align}
  \bw_{\mathrm{opt},q}\left(\hatbh_q\right) &= \frac{\bu_{1,q}}{\left\|\bu_{1,q}\right\|_2}.\label{eq:optbfvector}
\end{align}
\end{mylem}
%$E_{0,q} = \bbE_{\bh_q, \hatbh_q} \left[ \left|\bw^{H}(\hatbh_q) \bh_q \right|^2 \right] $

\begin{IEEEproof}
The harvested energy in one symbol period in~\eqref{EHEnergy1} can be rewritten as%in one symbol period
\begin{align}
E_0 &=\! \bbE_{\hatbh_q}  \!\left[ \bbE_{\bh_q | \hatbh_q} \! \left( \bw_q^{H} \! \left( \hatbh_q \!\right) \bh_q \bh_q^{H} \bw_q \! \left( \hatbh_q \! \right) \! \right) \! \right] \nonumber \\
&\!= \! \bbE_{\hatbh_q}  \! \left[ \! \bw_q^{H} \! \left( \hatbh_q \! \right)  \! \bR_{\bh_q | \hatbh_q} \bw_q \! \left(\hatbh_q \!\right) \! \right]. \label{totalenergymiso1}
\end{align}
%\bbE_{\bh_q, \hatbh_q} \!\left( \left|\bw^{H}(\hatbh_q) \bh_q \right|^2 \right)\!
Clearly, the $E_0$ is maximized when the beamformer is the largest eigenmode of $\bR_{\bh_q \left| \hatbh_q \right.}$, i.e., $\frac{\bu_{1,q}}{\left\|\bu_{1,q}\right\|_2}$.
\end{IEEEproof}
%$\bw_q(\hatbh_q)$
%The optimal beamformer in~\eqref{eq:optbfvector} is different from that in~\cite{MIMOWIPTZhang13}, since we consider the imperfect CSI.
%As stated in~\cite{MIMOWIPTZhang13}, the harvested energy is maximized when the energy beamforming the performed along the largest eigenmode of MIMO channel matrix, under the assumption %if perfect CSI.

%\begin{myrem}
Using the optimal beamformer in~\eqref{eq:optbfvector}, the normalized harvested energy is the mean of the largest eigenvalue $\gamma_1$ of the matrix $\bR_{\bh_q \left| \hatbh_q \right.}$. The total harvested energy in all remaining slots is thus given by
\begin{align}
 E &= (T-\tau) \; \bbE_{\hatbh_q} \left[ \gamma_1 \left(\bR_{\bh_q \left| \hatbh_q \right.} \right) \right], \label{totalenergymiso2}
\end{align}
where the matrix $\bR_{\bh_q \left| \hatbh_q \right.}$ is given by~\eqref{CondCorreMtx}.

We are interested in two widely-used channel estimators: least square (LS) and linear minimum mean-square-error (LMMSE) channel estimation~\cite{BigueshGershman06}. For both, we keep the transmit power for preamble a constant, which implies the effective receive signal-to-noise ratio (SNR) in the CE phase is only proportional to the preamble length $\tau$ that is to be optimized. We set the time duration for the preambles as $\tau= k m$, where $k=0,1,\cdots,N-1$. We next derive the optimal beamformer for the two estimators.

\subsection{Optimal Beamformer for LS Channel Estimation}
%Before giving the LS channel estimators, we from an information-theoretical perspective,
We first describe the optimal design of preamble. It is shown in~\cite{HassibiHochwaldIT03} that the LS estimation performance for both correlated and uncorrelated channels is optimized by using preamble $\bX$ that is a matrix with orthogonal columns, i.e., $\bX \bX^{H}= \frac{k}{m} \bI_{m}$, when total power for sending preambles is fixed as $k$. For simplicity, we choose the optimal preamble matrix as %follows that satisfies~\eqref{CovarianceX}
\begin{align}
  \bX = [\bX_1 \ \ \bX_2 \ \ \dots \ \bX_k]^T, \label{eq:PreambleMatrix1}
\end{align}
where $\bX_i= \frac{1}{\sqrt{m}} \bI_{m}$, for $i=1, 2, \cdots, k$. The transmit power for preamble is fixed as $\frac{1}{m}$. From~\cite{BigueshGershman06}, we obtain the LS estimate as follows
\begin{align}
\hatbh_{LS} &= \bX^{-1} \by = \bh + \frac{\sqrt{m}}{k} \sum \nolimits_{i=1}^{k} \bz_i,\label{eq:LSEstimate}
\end{align}
where the length-$m$ noise vector $\bz_i \sim \calC \calN (\mathbf{0}_m, \ \sigma_z^2  \bI_m)$. Clearly, the estimation error $\be$ is distributed as $\calC \calN \left( \mathbf{0}_m, \sigma_e^2 \bI_m\right)$, where $\sigma_e^2 = \frac{\sigma_z^2}{\beta} $ and $\beta = \frac{k}{m}=\frac{\tau}{m^2}$.
%%the $i$-th entry block of length-$m$ in the noise vector $\bz$
%$\sigma_e^2={m \sigma_z^2}/{k}$.
% \triangleq \hatbh - \bh$ follows complex Gaussian distribution $
%The covariance matrix of $\hatbh$ is thus $\bR_{\hatbh}=\bR+\bR_{\be}=\bR+\sigma_e^2 \bI_m$.

\subsubsection{Correlated channel}
%We use the training matrix in~\eqref{eq:PreambleMatrix1}.
%The general case of $q$-dimensional feedback is used.
From Lemma~\ref{mylem1}, we state that given $\hatbh_q$, the channel vector $\bh_q$ is distributed as $\calC \calN \left( \left(\sigma_e^2 \bR_q^{-1}+\bI_q \right)^{-1} \hatbh_q, \left(\bR_q^{-1}+\frac{1}{\sigma_e^2}\bI_q \right)^{-1} \right)$. The conditional correlation matrix $\bR_{\bh_q | \hatbh_q}$ yields
\begin{align}\label{eq:CovMatrixCorrelated}
\bR_{\bh_q | \hatbh_q} \!=\! &\left( \! \bR_q^{-1} \!+\! \frac{1}{\sigma_e^2}\bI_q \! \right)^{-1}  +  \left( \sigma_e^2 \bR_q^{-1} \!+\! \bI_q \right)^{-1} \hatbh_q \hatbh_q^{H} \left( \sigma_e^2 \bR_q^{-1} \!+\! \bI_q \right)^{-1}.
\end{align}
%where $\bR_q$ is the $q$-dimensional counterpart of channel covariance matrix $\bR$,
From~\eqref{eq:optbfvector}, the optimal beamforming vector is the largest eigenmode of $\bR_{\bh_q | \hatbh_q}$ in~\eqref{eq:CovMatrixCorrelated}.

\subsubsection{Uncorrelated channel}\label{OptBFUncorreLS}
From Lemma~\ref{mylem1}, we have that given $\hatbh_q$, the channel $\bh_q \sim \calC \calN \left( \frac{\hatbh_q}{1+ \sigma_e^2} , \frac{\sigma_e^2}{1+ \sigma_e^2} \bI_q \right)$. The conditional correlation matrix is thus rewritten as
\begin{align}\label{eq:CovMatrixUncorrelated}
\bR_{\bh_q \left| \hatbh_q\right.} &= \frac{\sigma_e^2}{1+ \sigma_e^2} \bI_q + \frac{\hatbh_q \hatbh_q^{H}}{\left(1+ \sigma_e^2 \right)^2}. %\label{CondCorreMtx}
\end{align}

Note that $\bR_{\bh_q \left| \hatbh_q \right.} $ is the sum of a scaled identity matrix and a rank-one matrix. The eigenvectors can be constructed as follows: take the normalized $\hatbh_q$ as the right eigenvector corresponding to the maximal eigenvalue, and construct other mutually orthogonal eigenvectors by Gram-Schmidt algorithm. Then from Lemma~\ref{LemmaOptBeamformer}, the optimal beamformer is thus given by
\begin{align}
  \bw_{\mathrm{opt},q}\left(\hatbh_q\right)= \frac{\hatbh_q}{\left\| \hatbh_q \right\|_2}.\label{eq:optbfvectorLSUncorre}
\end{align}
%Specially, if the channel estimation is perfect, i.e., $\hatbh_q=\bh_q$, then $\bw_{\mathrm{opt},q}=\bh_q^{H}/ \| \bh_q\|_2$. This is the same as the conventional maximum-ratio-transmit scheme.

Associated to the optimal beamformer in~\eqref{eq:optbfvectorLSUncorre}, the largest eigenvalue of the matrix $\bR_{\bh_q \left| \hatbh_q\right.}$ is
\begin{align}
\gamma_1\left(\bR_{\bh_q \left| \hatbh_q \right.}\right) &= \frac{\sigma_e^2}{1+ \sigma_e^2} + \frac{\left\| \hatbh_q \right\|_2^2}{\left(1+ \sigma_e^2\right)^2}. \label{eq:largesteigenvalue}
\end{align}
%= \left\| \hatbh_q \right\|_2^2 \left(k_1 + k_2\right)

\subsection{Optimal Beamformer for LMMSE Channel Estimation}
%\subsubsection{LMMSE Channel Estimation} \label{LMMSE_CE}
For an LMMSE estimator, the preamble sequences and per-antenna transmit power should be optimized by taking the spatial correlation of the channel into account~\cite{BigueshGershman06}.
%the training matrix in~\eqref{eq:PreambleMatrix1} is still optimal for spatially uncorrelated channel, but no longer optimal for correlated channel. When spatial correlation is present,
%optimal allocation of transmit power is in a water-filling manner. The
Denote the SVD of the channel covariance matrix as $\bR=\bB \bD \bB^{H}$. From~\cite{BigueshGershman06}, the optimal preamble matrix is
\begin{align}
  \bX = \sqrt{\sigma_z^2} \bF \left[ \left( \left[\mu_0 \bI_m - \bD^{-1}\right]^{+} \right)^{\frac{1}{2}}, \ \mathbf{0}_{m (k-1)}\right]^{T} \bB^H,\label{eq:PreambleMatrix2}
\end{align}
where the operator $[x]^{+}\triangleq \max \{0, x\}$, the $\bF$ is an arbitrary $km \times km$ unitary matrix, and the threshold $\mu_0$ is chosen subject to the power constraint\footnote{The transmit power for preamble is fixed as $\frac{1}{m}$.} $\tr \left({ \left[\mu_0 \bI_m - \bD^{-1} \right]^{+} } \right) = \frac{k}{\sigma_z^2}$.
%Without loss of generality, we choose $\bF$ as the normalized discrete Fourier transform matrix in the sequel, to allocate the preamble power in the whole CE phase.
% replaces all negative entries of a real matrix by zeros and leaves all nonnegative entries unchanged

Assuming the receive has perfect knowledge of $\bR$, from~\cite{BigueshGershman06}, the LMMSE channel estimate is given by
\begin{align}
  \hatbh_{LMMSE} = \bR \bX^{H} \left(\bX \bR \bX^{H} + \sigma_z^2 \bI_{\tau} \right)^{-1} \by
  \eqa \left(\sigma_z^2 \bR^{-1} + \bX^{H} \bX \right)^{-1} \bX^{H} \by,
  %&= \left(\bR^{-1} \sigma_z^2 + k \bI_m \right)^{-1} \bX^{H} \by,
  %\nonumber \\
\end{align}
where (a) follows from the well-known matrix inversion identity\footnote{Woodbury identity: Let $\bA$ and $\bB$ be positive definite matrix, then $\bA \bC^{H} (\bC \bA \bC^{H} + \bB)^{-1} = (\bA^{-1} + \bC^{H} \bB^{-1} \bC)^{-1} \bC^{H} \bB^{-1}$. }.
%The entries of the estimation error matrix $\be$ are known to be zero mean complex Gaussian random variables
It is straightforward to show that the error vector $\be$ is distributed as $\calC \calN (\mathbf{0}_m, \bR_{\be})$, where the error covariance matrix $\bR_{\be} =\left(\bR^{-1} + \frac{1}{\sigma_z^2}\bX^{H} \bX \right)^{-1}$. %\label{ErrorCovMatLMMSE}
%\end{align}
%The covariance matrix of $\hatbh$ is thus $\bR_{\hatbh}=\bR+\bR_{\be}=\bR+\left(\bR^{-1} + \bX^{H} \bX / \sigma_z^2 \right)^{-1}$.

\subsubsection{Correlated channel}
%, which implies $\bX^H \bX={k} \bI_m / {m}$ %We adopt the optimal preamble matrix in~\eqref{eq:PreambleMatrix2}.
From Lemma~\ref{mylem1}, the conditional correlation matrix $\bR_{\bh_q \left| \hatbh_q \right.}$ yields
\begin{align} \label{eq:CovMatrixCorrelatedLMMSE}
\bR_{\bh_q \left| \hatbh_q \right.}= \left( \bR_q^{-1} + \bR_{\be,q}^{-1} \right)^{-1}  + \left( \bR_{\be,q} \bR_q^{-1} + \bI_q \right)^{-1} \hatbh_q \hatbh_q^{H} \left( \bR_{\be,q} \bR_q^{-1} + \bI_q \right)^{-1}.
\end{align}
%where $\bR_q$ and $\bR_{\be,q}$ is the $q$-dimensional counterpart of channel covariance matrix $\bR$ and error covariance matrix $\bR_{\be} = \left(\bR^{-1} + \frac{k}{m \sigma_z^2} \bI_m \right)^{-1}$.
From~\eqref{eq:optbfvector}, the optimal beamforming vector is the largest eigenmode of $\bR_{\bh_q | \hatbh_q}$ in~\eqref{eq:CovMatrixCorrelatedLMMSE}.
%From~\eqref{eq:optbfvector}, the optimal beamforming vector is obtained as
%\begin{align}
%  \bw_{\mathrm{opt},q}\left(\hatbh_q\right)&= \frac{ \left( \bR_{\be,q} \bR_q^{-1} + \bI_q \right)^{-1} \hatbh_q}{\left\| \left( \bR_{\be,q} \bR_q^{-1} + \bI_q \right)^{-1} \hatbh_q \right\|_2}.\label{eq:optbfvectorLMMSECorre}
%\end{align}

\subsubsection{Uncorrelated channel}\label{OptBFUncorreLMMSE}
The optimal preamble matrix in~\eqref{eq:PreambleMatrix2} reduces to the orthogonal preamble matrix in~\eqref{eq:PreambleMatrix1}, when the channel is uncorrelated. Then we have $\bR=\bI_m$ and $\bR_{\be}=\frac{\sigma_z^2}{k+\sigma_z^2} \bI_m$.
%The optimal beamforming vector $\bw_{\mathrm{opt},q} (\hatbh_q) = \frac{\hatbh_q}{\left\| \hatbh_q \right\|_2}$, which is the same as~\eqref{eq:optbfvectorLSUncorre} for the case of LS channel estimation.
%We adopt the optimal preamble matrix in~\eqref{eq:PreambleMatrix2}.
From Lemma~\ref{mylem1}, the conditional correlation matrix $\bR_{\bh_q \left| \hatbh_q\right.}$ is given by
\begin{align}\label{eq:CovMatrixLMMSEUncorre}
\bR_{\bh_q \left| \hatbh_q \right.}
&= \frac{\sigma_e^2}{m + 2 \sigma_e^2} \bI_q + \left( \frac{m +  \sigma_e^2}{m + 2 \sigma_e^2} \right)^2 \hatbh_q \hatbh_q^{H},
\end{align}
which has similar structure to~\eqref{eq:CovMatrixUncorrelated}.
Following similar lines to the scenario that employs LS channel estimation in uncorrelated channels, the optimal beamforming vector is thus obtained from~\eqref{eq:optbfvector} as
\begin{align}\label{eq:optbfvectorLMMSEUncorre}
\bw_{\mathrm{opt},q} (\hatbh_q) = \frac{\hatbh_q}{\left\| \hatbh_q \right\|_2}.
\end{align}
%which is the same as~\eqref{eq:optbfvectorLSUncorre} for the case of LS channel estimation.
Associated to the optimal beamformer in~\eqref{eq:optbfvectorLMMSEUncorre}, the largest eigenvalue of the matrix $\bR_{\bh_q \left| \hatbh_q\right.}$ in~\eqref{eq:CovMatrixLMMSEUncorre} is
\begin{align}\label{LargestEigenvalueLMMSEUncorre}
  \gamma_1 \left( \bR_{\bh_q \left| \hatbh_q\right.} \right) =  \frac{\sigma_e^2}{m + 2 \sigma_e^2} + \left( \frac{m +  \sigma_e^2}{m + 2 \sigma_e^2} \right)^2 \left\| \hatbh_q \right\|_2^2.
\end{align}

\section{Wireless Power Transfer with Dynamic-length Preamble} \label{WPT_MISO_Dynamic}
In this section, we consider the scenario where the preamble length is allowed to vary dynamically depending on the current channel estimate. To maximize the expected harvested energy, we first formulate a dynamic programming (DP) problem~\cite{DPBertsekas05}, which will be shown to reduce to an optimal stopping problem, and thus can be simplified considerably. Using the optimal policy to the DP problem, we shall derive the optimal power allocation scheme in Section~\ref{DLPA}.

We assume uncorrelated flat Rayleigh fading channels which keep constant in each frame and vary independently among frames. The extension of the problem formulation for the more general case of Markovian channels is straightforward, see e.g.~\cite{HoOostveenLinnartz09}. Let $\hatbh_{k}$ denote the channel estimate available at time instant $k$. Assuming no priori channel knowledge is available, we initialize the channel estimate as the mean of $\bh$, i.e.,  $\hatbh_0=\mathbf{0}$. We assume that the receiver adopts an LS channel estimator and performs full (i.e., $m$-dimensional) feedback. The optimal beamformer in~\eqref{eq:optbfvectorLSUncorre} for $q=m$ is thus used in this section. We employ the preamble matrix in~\eqref{eq:PreambleMatrix1}. For $\ k=0,\ 1, \cdots, N-2$, it is useful to rewrite the LS channel estimate in~\eqref{eq:LSEstimate} as the following recursive equation
\begin{align} \label{eq:hkrecursive}
\hatbh_{k+1}  = \bh + \frac{\sqrt{m}}{k+1} \sum \nolimits_{i=1}^{k+1} \bz_i =\frac{k}{k+1} \hatbh_k + \frac{\bh}{k+1} + \frac{\sqrt{m} \bz_{k+1}}{k+1}.
\end{align}%convenient
%For simplicity, w

Before formulating the problem and obtaining the solutions, we first obtain useful statistical properties on the channel estimates.

\subsection{Statistical Properties of Channel Estimates}\label{Section-IV-A}
Lemma~\ref{LemmaRecursiveCondDist} will quantity the statistical relationship of two channel estimates at adjacent time instants; while Lemma~\ref{LemmaFirstOrderMarkov} will show that the most recent channel estimate provides sufficient statistics for estimating the channel.
\begin{mylem}\label{LemmaRecursiveCondDist}
Given $\hatbh_k$, the next channel estimate $\hatbh_{k+1}$ is distributed as $\calC \calN (\bar{\bu}_k, \bar{\sigma}_k^2 \bI_m)$, where
\begin{align}
  \bar{\bu}_k = \frac{k(k+1+m \sigma_z^2)}{(k+1)(k+m \sigma_z^2)}\hatbh_k, \quad  \bar{\sigma}_k^2 = \frac{m \sigma_z^2 (k+1 + m \sigma_z^2)}{(k+1)^2 (k+m \sigma_z^2)}. \nonumber
\end{align}
\end{mylem}

\begin{IEEEproof}
Let $\sigma_k^2=\frac{m}{k} \sigma_z^2$. From Lemma~\ref{mylem1} for $q=m$, we have that $\bh \left| \hatbh_k \right. \sim \calC \calN \left(\frac{1}{1+\sigma_k^2} \hatbh_k, \frac{\sigma_k^2}{1+\sigma_k^2} \bI_m\right)$. From~\eqref{eq:hkrecursive} and the independence between $\hatbh_k$ and the noise vector $\bz_{k+1}$, we obtain the result after algebraic manipulations.
\end{IEEEproof}

%Preparatively, we first derive the pdf of $H_k$. Using the optimal policy $\pi^{\star}$, we denote the joint pdf of $H_1,H_2,\cdots,H_k$ by
%Since the real part and the imaginal part of each element in $\hatbh_{k+1} $ are independent,
We take the channel estimate power as a random variable $V_k$, i.e., $V_k \triangleq \| \hatbh_k \|_2^2$. From Lemma~\ref{LemmaRecursiveCondDist}, conditioned on $V_k=v_{k}$, the random variable $\frac{2} {\bar{\sigma}_k^2} V_{k+1}$ follows the noncentral Chi-Square distribution with the degree $2m$ of freedom and the noncentrality parameter
\[\theta_k= \frac{2 k^2 (k+1+m \sigma_z^2)v_k } {m \sigma_z^2 (k+ m \sigma_z^2)}.
\]
%=\sum_{i=1}^m \left( \frac{2} {\bar{\sigma}_k^2} \left\| \mathrm{Re} \left(\bu_k\right) \right\|_2^2 + \frac{2} {\bar{\sigma}_k^2} \left\| \mathrm{Im} \left(\bu_k\right) \right\|_2^2 \right)
Moreover, the conditional probability density function (pdf) of $V_{k+1}$ is thus given by~\cite{JGProakisMSalehi05}%, denoted by $f(v_{k+1} | v_k)$,
\begin{align}
f(v_{k+1} | v_k) = &\frac{1}{\bar{\sigma}_k^2} \exp{ \left(- \frac{v_{k+1}}{\bar{\sigma}_k^2}-\frac{\theta_k}{2}\right)} \left( \frac{2v_{k+1}} {\theta_k \bar{\sigma}_k^2}\right)^{\frac{m-1}{2}} I_{m-1} \left( \frac{2 \theta_k v_{k+1} }{\bar{\sigma}_k^2}\right),\label{eq:cond_pdf}
\end{align}
where $I_{m\!-\!1}(\cdot)$ is the $(m\!-\!1)$-th order modified Bessel function of the first kind. The conditional mean is
\begin{align}\label{eq:Mean_NoncentralChiSquare}
\bbE_{V_{k+1} | V_k = v_k}  \left( V_{k+1} \right) = \bar{\sigma}_k^2 \left(m+\frac{\theta_k}{2}\right).
\end{align}
%, which is a linear function of the previous channel estimate power $v_k$.  %\label{eq:Mean_NoncentralChiSquare}
%\end{align}
%%, and the cumulative distribution function of $\| \hatbh_{k+1} \|_2^2 |_{\hatbh_k}$ is given by
%\begin{align}
%F(h)=1-Q_m \left( \sqrt{\theta_k}, \; \sqrt{\frac{2 h}{\sigma_k^2}} \right),\label{eq:cond_cdf}
%\end{align}
%where $Q_m(\cdot)$ is the $m$-th Marcum Q-function

\begin{mylem}\label{LemmaFirstOrderMarkov}
Given a sequence of LS channel estimates $\hatbh_1, \hatbh_2, \cdots,\hatbh_k$, the distribution of channel vector $\bh$ conditioned on all channel estimates is simplified as
\begin{align}
    f \left(\bh \left| \hatbh_1, \hatbh_2, \cdots, \hatbh_{k}\right. \right)=f \left(\bh \left| \hatbh_{k} \right. \right), %\nonumber
\end{align}
which is the Gaussian distribution $\calC \calN \left(\frac{1}{1+\sigma_k^2} \hatbh_k, \frac{\sigma_k^2}{1+\sigma_k^2} \bI_m\right)$, where $\sigma_k^2=\frac{m}{k} \sigma_z^2$. %with covariance matrix given by~\eqref{eq:CovMatrixCorrelated} for correlated channel and by~\eqref{eq:CovMatrixUncorrelated} for uncorrelated channel, where $q=m$, respectively.
\end{mylem}

\begin{IEEEproof}
  See proof in Appendix~\ref{AppFirstOrderMarkov}.
\end{IEEEproof}
Lemma~\ref{LemmaFirstOrderMarkov} suggests that the accuracy of channel estimation can not be increased by using all available channel estimates, compared to using only the most recent channel estimate. This observation will be used to show the structure of the optimal DP policy (see Lemma~\ref{LemOtpStopping}, later).
%define the system state as well as

\subsection{Problem Formlation}% Dynamic Programming Formulation}
We formulate the optimization problem to maximize the total expected harvested energy, assuming the transmitter uses constant transmit power for WPT. %The optimal solution will be obtained via dynamic programming using Bellman's equation \cite{DPBertsekas05}.
We first make the necessary definitions.
Consider slot $k$, where $k=0, \ 1, \ \cdots, \ N-1$.
\subsubsection{Decision (or Control) Variable}
We denote the decision  variable as $u_k\in \calC$. The decision space $\calC$ consists of only two elements $\mathsf{s}$ and $\mathsf{c}$, that corresponds to stopping CE (i.e., requesting WP) or continuing CE, respectively. We initialize $u_{-1}=\mathsf{c}$.
\subsubsection{System State}
We define the system state $\bx_k$ as consisting of (i) $\delta_k$ which denotes the number of slots used so far for CE, and (ii) the most recently available channel estimate.
%\footnote{\textcolor{red}{To check} Having all the past channel estimates available does not improve the amount of expected harvested energy, compared to having the most recent channel estimate. This is because $p(\bh|\bh_k, \bh_{k-\tau_1}, \bh_{k-\tau_2},\cdots)=p(\bh|\bh_k)$ for $\tau_i>0$, i.e., the conditional pdf of the channel depends only on the most recent channel estimate.}.
Given $u_k$ and current state $\bx_{k}$, the next state is
 %and the slot when the channel estimate current decision $u_k$ and the  =\{u_k, \hatbh_k\}$ with state space $\calS= \left\{ \calC, \bbC^{m} \right\}$. Given $\bx_k$, the system state thus transits to the next slot according to
\begin{equation}\label{eq:state}%\forall 1 \leq i  \leq k, \;\;
\bx_{k+1}=\left\{ \begin{array}{cl}
\{\delta_k+1, \hatbh_{k+1}\}, \quad &{\text{if}} \;\; u_k = \mathsf{c} \\
\bx_{k}, \quad &{\text{if}} \;\; u_k = \mathsf{s}.
\end{array} \right.
\end{equation}
The initial state is $\bx_0 =\{\delta_0,\hatbh_0 \}$ with $\delta_0=0, \hatbh_0=\emptyset$. We denote the space of all possible state as $\calS$. From Lemma~\ref{LemmaFirstOrderMarkov}, this system state is sufficient to obtain the statistics of $\bh$ even if all post channel estimates were made available.
%using all the past channel estimates does not improve the harvested energy.
%By writing that the system is at a state $\bx_{k+1}= \left\{u_k, \; \hatbh_{k+1}\right\}$, we mean that the transmitter-receiver pair performs channel estimation in the previous $(k+1)$ slots, and the available channel estimation at the beginning of the $(k+1)$-th period is $\hatbh_{k+1}$.
\subsubsection{Policy}
Define a policy $\pi$ as a sequence of functions
\[\pi = \{\mu_k(\bx_k), \; \forall \bx_k \in \calS, k=0,1,\cdots,N-1\},
\]
where $\mu_k:\calS\rightarrow\calC$ is a function that maps the state $\bx_k$ into the decision variable in the next time slot, i.e., $u_k = \mu_k(\bx_k)$. We denote the set of all possible policies as $\mathit{\Pi}$.

\subsubsection{Reward}
Given state $\bx_k$ and decision $u_k= \mu_k(\bx_k)$, we denote $g_k(\bx_k, u_k) $ as the reward, given by the expected harvested energy in slot $k$. If $u_k=\mathsf{s}$, we have from %\eqref{PerSlotenergymisoLSUncorre}
\eqref{eq:largesteigenvalue} with $\sigma_e^2=\frac{m}{\delta}\sigma_z^2$ that
\begin{align}\label{eq:rewar_in_one_slot}
  g_k(\bx_k, u_k) %&= m \cdot \left(\bw_{\mathrm{opt}}(\bx_k)^{H}  \bR_{\bh_q | \bx_k} \bw_{\mathrm{opt}}(\bx_k) \right) \nonumber \\
  &=  m \cdot \left( \frac{m \sigma_z^2}{\delta+ m \sigma_z^2} + \frac{\delta^2 \left\| \bh_k \right\|_2^2 }{(\delta+m \sigma_z^2)^2} \right),
\end{align}
and  if $u_k=\mathsf{c}$, then $g_k(\bx_k, u_k)=0$.
%the harvested energy in one time slot can be obtained
\subsubsection{Dynamic Program and Optimal Policy}
%We are now ready to state the problem.
To maximize the total harvested energy, we thus have the optimization problem $\mathrm{P_1}$ given in~\eqref{eq:optimDP}. With the problem formulation $\mathrm{P_1}$, the optimal policy $\pi^{\star}$ is given by the functions $\{\mu_k(\cdot)\}$, i.e., the decision $u_k$ given state $\bx_k$,  that satisfy the Bellman's equation~\cite{DPBertsekas05}:
\begin{align}
%  J_K(\bx_K) &= E(\hatbh_K, K), \nonumber \\
  &J_{N-1}(\bx_{N-1}) \!=\! \max_{u_{N-1}} \; g_{N-1}(\bx_{N-1}, u_{N-1}),\nonumber \\
   &J_{k}(\bx_{k}) \!=\! \max_{u_{k}} g_k(\bx_k, u_k) \!+\! \bbE_{\hatbh_{k+1}| {\hatbh_k} } \left[ J_{k+1}(\bx_{k+1})\right],
\label{eq:Bellman}
\end{align}
for $k=N-2, \cdots, 0$. Here, $J_{k}(\bx_{k})$ is known as the value function which represents the harvested energy for the last $(N-k)$ time slots, conditioned on the current system state $\bx_{k}$.
Typically, the solution is obtained by backward recursion, by first solving for $\mu_{N-1}(\cdot)$ for slot $N-1$, then for $\mu_{N-2}(\cdot), \cdots, \mu_0(\cdot)$.

\subsection{Optimal Policy} \label{OptPolicy}
Lemma~\ref{LemOtpStopping} states that the Bellman's equation \eqref{eq:Bellman} can be reduced to an optimal stopping problem, for which a decision is changed at most once and fixed henceforth.

\begin{mylem}\label{LemOtpStopping}
%Assuming the uncorrelated channel is estimated by an LS estimator.
Any decision sequence of the optimal policy $\pi^{\star}$ has the structure
%that $u^*_k=\mathsf{c}$ for $0\leq k \leq k^*$ and $u^*_k=\mathsf{s}$ for $k^*+1 \leq k \leq N-1$, where
\begin{align}
 (u^*_0, u^*_1,\cdots, u^*_{N-1}) = (\underbrace{\mathsf{c}, \mathsf{c}, \cdots, \mathsf{c}}_{k=0,1,\cdots \!,k^{\ast}-1}, \underbrace{\mathsf{s},\mathsf{s},\cdots,\mathsf{s}}_{k=k^{\ast},\cdots,N-1}), \end{align}
where $0\leq k^{\ast}\leq N-1$. That is, the optimal policy initially performs only CE for the first $k^*$ slots, then performs only WP for the remaining slots.
\end{mylem}

\begin{IEEEproof}
See proof in Appendix~\ref{AppOptStopping}.
\end{IEEEproof}
Lemma~\ref{LemOtpStopping} allows us to simplify the DP problem and obtain a solution that can be implemented with low complexity. Before we obtain the statement of the optimal policy in Theorem~\ref{TheoremThreshold}, we first state the expected harvested energy under different scenarios. Henceforth, we assume the optimal policy is employed.
%giving the optimal policy, we define the conditional harvested energy.

Given state $\bx_k$, if the receiver decides to request for WP, i.e., $u_k = \mathsf{s}$, then the expected harvested energy in the remaining slots is obtained from~\eqref{totalenergymiso2} and~\eqref{eq:largesteigenvalue} as%~\eqref{eq:optbfvector} and~\eqref{totalenergymiso1} as
\begin{align}%\bE_{\text{cond}} \triangleq
 \tilE \left(\hatbh_k, k \right) = m(N-k)  \left[ \bbE_{\bh \left| \hatbh\right.} \left( \bw_{k,{\text{opt}}}^{H} \bh \bh^{H} \bw_{k,{\text{opt}}} \right) \right] = A_k \left( B_k + C_k \left\| \hatbh_k \right\|_2^2\right), \label{eq:CondEnergy}
%  = m(N-k) \left( \frac{m \sigma_z^2}{k + m \sigma_z^2} + \frac{k^2 \| \hatbh_k \|_2^2 }{(k+m \sigma_z^2)^2} \right).
\end{align}
where $A_k=m(N- k)$, $B_k=\frac{m \sigma_z^2}{k + m \sigma_z^2}$, and $C_k=\frac{k^2}{(k+m \sigma_z^2)^2}$. If the decision is instead $u_k=\mathsf{c}$, then the expected harvested energy in the last $(N-k)$ time slots, under all possible decisions made for subsequent slots, is given by
\begin{align}\label{CondharvestedEnergy}
{\bar{J}}_{k+1} \left(\bx_{k}\right) &= \bbE_{\hatbh_{k+1} \left| {\hatbh_k} \right.} \left[ J_{k+1}\left(\bx_{k+1}\right) \right].
\end{align}

For the special case in which the receiver decides to continue CE ($u_k=\mathsf{c}$) at time instant $k$ and stop CE ($u_{k+1}=\mathsf{s}$) at time instant $k+1$, then from the conditional mean in~\eqref{eq:Mean_NoncentralChiSquare}, the expected harvested energy conditioned on $\hatbh_k$ is obtained after some algebraic manipulation as
\begin{align}\label{CondharvestedEnergySpecialCase}
{\barE} \left(\hatbh_{k+1}, k \!+\! 1\right) = \bbE_{\hatbh_{k+1} \left| {\hatbh_k} \right.} \left[\tilE \left(\hatbh_{k+1}, k \!+\! 1 \right) \right]  =D_{k+1} \frac{k^2 (k \!+\! 1 \!+\! m \sigma_z^2) \left\| \hatbh_{k} \right\|_2^2 \!+\! F_{k+1}}{G_{k+1}}, % \nonumber  %\label{eq:condenergy_stop_nextinstant} \nonumber \\
\end{align}
where $D_{k+1} =m(N\!-\!k\!-\!1),\ F_{k+1} =m \sigma_z^2 (k\!+\!m \sigma_z^2) (k\!+\!m \sigma_z^2 \!+\! m)$, and $G_{k+1} =(k\!+\!1\!+\!m \sigma_z^2) (k+m \sigma_z^2)^2$.

Now, we state the optimal policy in Theorem~\ref{TheoremThreshold}.

\begin{mythe}\label{TheoremThreshold}
The optimal policy to Problem $\mathrm{P_1}$ is a threshold type policy that depends only on the channel estimate power, i.e.,
\begin{align}\label{eq:Multiple_threshold}
  u_k&= \left\{ \begin{array}{cl}
    \mathsf{c}, \quad &{\textrm{if}} \ \ \left\| \hatbh_k \right\|_2^2 \in \calD_{\mathsf{c},k}   \\ %\nonumber
        \mathsf{s}, \quad &{\textrm{if}} \ \ \left\| \hatbh_k \right\|_2^2 \in \calD_{\mathsf{s},k}
  \end{array} \right.   %\label{eq:Multiple_threshold}
\end{align}
where the sets (intervals)
\begin{align}
  \calD_{\mathsf{c},k}=\big[[0, \lambda_{k,1}) \cup [\lambda_{k,2}, \lambda_{k,3}) \cup \cdots \cup [\lambda_{k,M_k-1}, \lambda_{k,M_k})\big], \nonumber \\
  \calD_{\mathsf{s},k} =\big[[\lambda_{k,1}, \lambda_{k,2}) \cup [\lambda_{k,3}, \lambda_{k,4}) \cup \cdots \cup [\lambda_{k,M_k},+\infty)\big], \nonumber
\end{align}
and $\lambda_{k,1} \leq \cdots \leq \lambda_{k,M_k}$ are the solutions to $\tilE(\hatbh_{k}, k)={\bar{J}}_{k+1} (\bx_{k+1})$ with respect to the variable $\| \hatbh_k \|_2^2$.
\end{mythe}
%$\{\lambda_{k,j}, \; j=1,\cdots, M_k\}$,

\begin{IEEEproof}%{DPBellman72}
From Lemma~\ref{LemOtpStopping} and~\cite{DPBertsekas05}, to obtain the optimal policy for the original DP problem, if $u_{k-1}=\mathsf{s}$, then $u_{k}=\mathsf{s}$; if $u_{k-1}=\mathsf{c}$, it suffices to solve the recursive Bellman's equation
\begin{align}
     J_{k}(\bx_{k}) %\nonumber \\
     &= \max \left\{ \tilE \left(\hatbh_{k}, k \right), {\bar{J}}_{k+1} (\bx_{k} ) \right\} \label{eq:DP_algorithm1}
%     \left\{\begin{array}{cl}
%  \max \left\{ \tilE(\hatbh_{k}, k), {\bar{J}}_{k+1} (\bx_{k} ) \right\}, \; &{\textrm{if}} \; u_{k-1}=\mathsf{c} \\
%   0, \; &{\textrm{if}} \; u_{k-1}=\mathsf{s}
%   \end{array} \right.  \label{eq:DP_algorithm1}
\end{align}
for $k=N-2, N-1, \cdots, 0$ in a backward manner. Specifically, assuming $u_{k-1}=\mathsf{c}$, if $\tilE(\hatbh_{k}, k) > {\bar{J}}_{k+1} (\bx_{k} )$, then the receiver requests WP with $u_{k}=\mathsf{s}$, otherwise the receiver continues CE with $u_{k}=\mathsf{c}$. Hence, to obtain the optimal decision, we have to solve the equation % by comparing the expected harvested energy
\begin{align}
 &\tilE \left(\hatbh_{k}, k \right) ={\bar{J}}_{k+1} (\bx_{k}) \label{eq:threshold_equation}
%  &=\bbE_{\hatbh_{k+1} } \left[ \max \left\{ E(\hatbh_{k+1}, k+1), {\bar{J}}_{k+2} (\bx_{k+2} | {\hatbh_{k+1}}  )  \right\} | {\hatbh_k}  \right] \nonumber
\end{align}%in a backward manner,
to compare the two terms in~\eqref{eq:DP_algorithm1}. From~\eqref{eq:CondEnergy}, the first term $\tilE (\hatbh_{k}, k)$ is a monotonic increasing linear function of only the channel estimate power $\| \hatbh_{k} \|_2^2$. Moreover, the second term ${\bar{J}}_{k+1} (\bx_{k})$, given in~\eqref{CondharvestedEnergy}, is also a function of only $\| \hatbh_{k} \|_2^2$. This claim is checked by induction based on backward recursion as follows. At time instant $k=N-2$, we have ${\bar{J}}_{N-1} (\bx_{N-2}) ={\barE} (\hatbh_{N-1}, N-1)$. From~\eqref{CondharvestedEnergySpecialCase}, the term ${\bar{J}}_{N-1} (\bx_{N-2})$ is a linear function of $\| \hatbh_{N-2} \|_2^2$. It follows from~\eqref{eq:DP_algorithm1} that $J_{N-2} (\bx_{N-2})$ is a piecewise linear function of $\| \hatbh_{N-2} \|_2^2$. From~\eqref{CondharvestedEnergy} and the conditional distribution in Lemma~\ref{LemmaRecursiveCondDist}, we thus obtain that $\bar{J}_{N-2} (\bx_{N-3})$ is a function of $\| \hatbh_{N-3} \|_2^2$. By mathematical induction with decreasing slot index, we have that $\bar{J}_{k+1} (\bx_{k})$ is a function of $\| \hatbh_{k} \|_2^2$.

Hence, the decision policy is threshold-type possibly with multiple thresholds for $\| \hatbh_{k} \|_2^2$. Denote the solution(s) to~\eqref{eq:threshold_equation} with respect to $\| \hatbh_{k} \|_2^2$ by $\lambda_{k,1}, \; \lambda_{k,2}, \cdots,\lambda_{k,M_k}$, assuming $\lambda_{k,1} \leq \cdots \leq \lambda_{k,M_k}$. The desired result is obtained.
\end{IEEEproof}

In general, the state value is of $(m+1)$-dimension and the complexity of implementing the policy can be very high. Moreover, the computational complexity to obtain the optimal policy can be prohibitive and the memory required to store the policy too large.
From Theorem~\ref{TheoremThreshold}, however, the optimal policy can be implemented for each slot by only comparing the channel estimate power to a scalar value, thus saving complexity in computation and storage of policy. The thresholds can be pre-computed and stored in a lookup table. During online implementation, the receiver refers to the table to make the decision.

\subsection{Optimal Thresholds}
In this section, we derive the optimal thresholds $\{\lambda_{k,j}\}$ in Theorem~\ref{TheoremThreshold} in a backward manner, by solving equation~\eqref{eq:threshold_equation} for $k=N-1, N-2, \cdots,0$, assuming $u_{k-1}=\mathsf{c}$. At time instant $N-1$, we have $\tilE (\hatbh_{N-1}, N-1)>0, \; \forall \ \hatbh_{N-1} \in \bbC^{m}$. Thus, it is optimal to set $\lambda_{N-1}=0$. This is because it is always optimal for the receiver to stop CE at time instant $N-1$, since $\| \hatbh_{N-1} \|_2^2 >0$ holds with probability one for Rayleigh fading channels.
%the threshold

At time instant $k=N-2$, we have ${\bar{J}}_{N-1} (\bx_{N-2}) ={\barE} (\hatbh_{N-1}, N-1)$. The equation~\eqref{eq:threshold_equation} for $k=N-2$ thus reduces to $\tilE (\hatbh_{N-2}, N-2)={\bar{E}} (\hatbh_{N-1}, N-1)$. Observe that the left-hand side (LHS) and right-hand side (RHS) are both monotonically increasing linear functions of $\| \hatbh_{N-2} \|_2^2$. Hence, the decision policy at this time instant is a threshold-type with a single threshold\footnote{Even if the two linear functions are parallel, by approximate construction, we can still use a single threshold without loss of generality.}. If $D_{N-1} (N-2)^2 \neq A_{N-2} G_{N-1} C_{N-2}$, the threshold is
\begin{align}
  \lambda_{N-2} \!&=\! \left[ \frac{A_{N-2} B_{N-2} G_{N-1}-D_{N-1} F_{N-1}}{D_{N-1} (N-2)^2 - A_{N-2} G_{N-1} C_{N-2}} \right]^{+},
\end{align}
where the notation $[x]^{+} = \max\{0, x\}$; and $\lambda_{N-2}=0$, otherwise.
%, by setting an approximate threshold taking into consideration if there are multiple or no solution(s) for the equation
For the subsequent slots $k=N-3,N-4,\cdots,0$, we can obtain the thresholds by a numerical search as follows. By substituting~\eqref{eq:DP_algorithm1} into~\eqref{CondharvestedEnergy}, we note that the RHS of~\eqref{eq:threshold_equation} is expressed in a recursive form
\begin{align}
  {\bar{J}}_{k+1} (\bx_{k}) &=  \bbE_{\hatbh_{k+1} \left| {\hatbh_k} \right. } \left[ \max \left\{ \tilE \left(\! \hatbh_{k+1}, k \!+\! 1 \!\right), {\bar{J}}_{k+2} (\bx_{k+1} ) \right\} \right]. \label{eq:threshold_equation_expanded} %={\bar{J}}_{k+1} (\bx_{k} )
\end{align}
%\tilE \left(\hatbh_{k}, k\right) it is not an explicit expression.
Hence, it is difficult to obtain a closed-form solution of the threshold $\| \hatbh_{k} \|_2^2$ that solves~\eqref{eq:threshold_equation}. Thus, we let the quantity $\| \hatbh_{k} \|_2^2$ take discrete values in the set $\calQ \triangleq \{\Delta,\; 2\Delta, \; \cdots, \;M\Delta \}$, and search for the closest values(s) in the set $\calQ$ that solves~\eqref{eq:threshold_equation}.

%Fig.~\ref{fig:FigMultipleSolutions} illustrates the LHS and RHS of~\eqref{eq:threshold_equation}, in which two possible functions for the RHS are plotted.
In general, there may be multiple solutions to~\eqref{eq:threshold_equation}, denoted as $\lambda_{k,j}, \; j=1,\cdots, M_k$, since the LHS $\tilE(\hatbh_{k}, k)$ is a monotonic increasing linear function of $\| \hatbh_{k} \|_2^2$, and the RHS is a function of $\| \hatbh_{k} \|_2^2$. To get more insights, we give a numerical example on the thresholds.
%\begin{figure}[t]%htbp
%\centering
%\includegraphics[width=.99\columnwidth] {figs/MultipleSolutions1013.eps}
%%\caption{Plot of MSE v.s. the length of preamble.}
%\caption{Illustration of general multi-threshold policy.}
%\label{fig:FigMultipleSolutions}
%\end{figure}

%In the following, we give a numerical example on the thresholds.
\begin{myexa}\label{ExampleThreshold}
Let $T=126, m=3,\sigma_z^2 = 1$. The threshold is numerically computed and shown in Fig.~\ref{fig:Fig6}. We numerically find that the threshold at each time index $k$ is always unique, which can further simplify the decision process in practice.
\begin{figure}[t]
\centering
\includegraphics[width=.7\columnwidth] {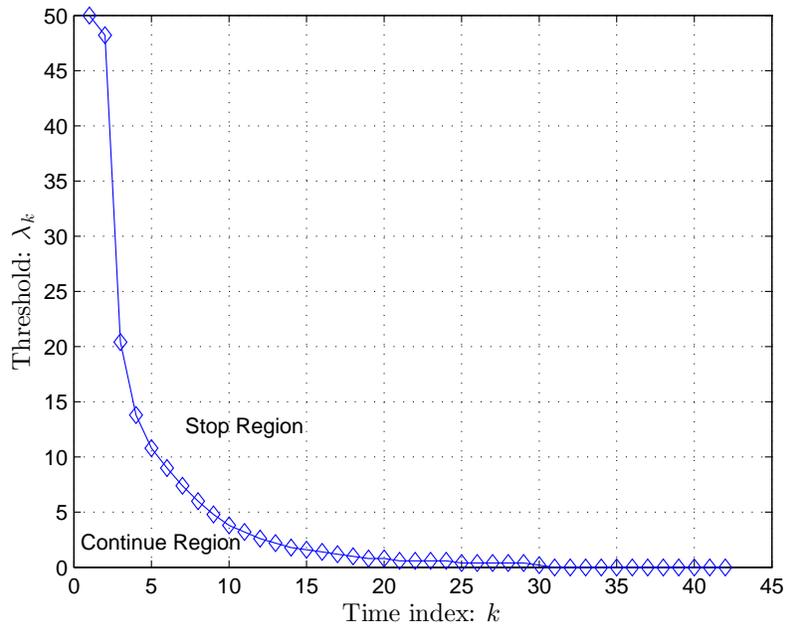}
%\caption{Plot of MSE v.s. the length of preamble.}
\caption{Threshold over slot index $k$.}
\label{fig:Fig6}
\end{figure}
It is observed that the threshold monotonically decreases as the time index $k$ increases. This observation is consistent with the intuition that if a channel estimate is good enough to be acceptable at time $k$ for WP to be performed, it should also be acceptable at time $(k+1)$ when there will be one more slot available for the channel estimate to be improved.
\end{myexa}

\section{Wireless Power Transfer with Fixed-length Preamble} \label{WPT_MISO_Static}
Based on the optimal beamformer in~\eqref{eq:optbfvector}, in this section, we consider the scenario wherein the preamble length is fixed in all frames. We first derive the optimal preamble length. This corresponds to the case of offline adaptation, in contrast to online adaptation in Section~\ref{WPT_MISO_Dynamic} where the preamble length is varied over frames. Then, we derive the optimal scheme for allocating power according to the channel estimate power.%different
%performed

\subsection{WPT with LS Channel Estimator}\label{OptFixedLength_LS}
\subsubsection{Correlated channel}
From~\eqref{totalenergymiso2} and the conditional correlation matrix in~\eqref{eq:CovMatrixCorrelated}, the harvested energy is
\begin{align}
 E &= (T-\tau)  \bbE_{\hatbh_q} \left[ \gamma_1 \left( \left(\bR_q^{-1}+ \frac{1}{\sigma_e^2}\bI_q \right)^{-1} + \left( \sigma_e^2 \bR_q^{-1} + \bI_q \right)^{-1} \hatbh_q \hatbh_q^{H} \left( \sigma_e^2 \bR_q^{-1} + \bI_q \right)^{-1} \right) \right]. \label{totalenergymisoLSCorre}
\end{align}
For the general correlated channel, it appears that there is no closed-form expression. Thus, it is difficult to analytically obtain the optimal length of preamble. However, by an one-dimensional search, we can numerically find the optimal length of preamble, which shall be shown in Section~\ref{Section_NumericalResults}.
%the expectation can be numerically obtained. B
% for the expectation in~\eqref{totalenergymisoLSCorre}
\subsubsection{Uncorrelated channel}
Using the optimal beamformer in~\eqref{eq:optbfvectorLSUncorre}, from~\eqref{totalenergymiso2} and~\eqref{eq:largesteigenvalue}, the total harvested energy is rewritten as%with optimal beamformer
\begin{align}
 E %&= (T \!-\!\tau)  \bbE_{\hatbh_q}  \left[ \bw_{\mathrm{opt},q}^{H}  \bR_{\bh_q \left| \hatbh_q \right.} \bw_{\mathrm{opt},q} \right]  \nonumber \\
 &= (T \!-\! \tau)   \left[ \! \frac{\sigma_e^2}{1+ \sigma_e^2} + \frac{1}{(1+ \sigma_e^2)^2} \bbE_{\hatbh_q} \left( \left\| \hatbh_q \right\|_2^2\right) \! \right]. \label{totalenergymisoLSUncorre}
\end{align}

Before giving the result, we define a quantity that depends on only the number of transmit antennas $m$ and the feedback dimension $q$ as
\begin{align}
      G_{m,q} &\triangleq  \sum \nolimits_{r=1}^q  \frac{2 m!}{(r-1)!}  \sum \nolimits_{s=1}^{m-r+1} \frac{s (-1)^{s+1}}{(m\!-\!r\!+\!1\!-\!s)! s! (s\!+\!r\!-\!1)^2}. \label{eq:G_mqConstant}
\end{align}%\sum \nolimits_{r=1}^q g_{m,r}     =
It can be shown that $G_{m,q}$ increases as either $m$ or $q$ increases. In the case of full feedback, i.e., the receiver feeds back $\hatbh$ to the transmitter, we have $G_{m,m}=2m$.
%As seen in Appendix~\ref{AppendixC}, we observe that $G_{m,q}$ increases as either $m$ or $q$ increases. In the case of full feedback, i.e., the receiver feeds back $\hatbh$ to the transmitter, we have $G_{m,m}=2m$.
%(i.e., the $m$-dimensional feedback $\hatbh_m$)

In independent Rayleigh-fading MISO channels, by using the optimal beamformer in~\eqref{eq:optbfvectorLSUncorre} at the transmitter, we have the following theorem.
\begin{mythe} \label{mythe1}
Let $T, m,\sigma_z^2 $ and $G_{m,q}$ be defined as before. When the channel is estimated by an LS estimator, the {\textit{unique}} optimal length of preambles for channel estimation is given by
\begin{equation} \label{eq:OPtLength}
\tau^{\star} = \left\{ \begin{array}{cl}
  0, \;\; &{\text{if}} \;\;
  \sigma_z^2 > \frac{T(G_{m,q} - 2)}{2 m^2}  \\
  \arg \underset{\tau \in \{\lfloor \tau_1 \rfloor, \lceil \tau_1 \rceil\} }{\max} E \left(\tau\right), \;\; &{\text{otherwise}}
 \end{array} \right.
\end{equation}
%where the quantity $
%  \tau_1 = - m^2 \sigma_z^2 + m \sqrt{ {\sigma_z^2}(m^2 \sigma_z^2 + T)(G_{m,q}-2) / {G_{m,q}} }$, the notations $\lfloor \cdot \rfloor$ and $\lceil \cdot \rceil$ are the floor operation and the ceiling operation, respectively, and the function
where the notations $\lfloor \cdot \rfloor$ and $\lceil \cdot \rceil$ are the floor operation and the ceiling operation, respectively, the quantity
\[
  \tau_1 = - m^2 \sigma_z^2 + m \sqrt{ \frac{{\sigma_z^2}(m^2 \sigma_z^2 + T)(G_{m,q}-2)} { {G_{m,q}} }}
  \]
and the function
\begin{align}
 E \left(\tau\right) = (T-\tau)  \frac{G_{m,q} \tau + 2 m^2 \sigma_z^2}{2 \left(\tau+m^2 \sigma_z^2\right)}, \label{Energy_given_tau_L}
\end{align}
Moreover, the corresponding maximal harvested energy $E_{\max}=E \left( \tau^{\star} \right)$.
%\begin{equation} \label{eq:maxenergy}
%E_{\max} =  \left\{ \begin{array}{cl}
%  T, \;\; &{\text{if}} \;\; T <  \frac{2 m^2 \sigma_z^2}{G_{m,q} - 2}  \\
%  E (\tilde{\tau}), \;\; &{\text{otherwise}}.
% \end{array} \right.
%\end{equation}
\end{mythe}

\begin{IEEEproof}
See proof in Appendix~\ref{AppendixB}.
\end{IEEEproof}
We see that the quantity $G_{m,q}$ in the maximal harvested energy $E_{\max}$ represents the effect of $q$-dimensional (partial) feedback on the harvested energy.

\begin{myrem}[Optimization over the number of transmit antennas] \label{remark1}
In this paper, we assume the number of transmit antennas is fixed as $m$. Note that using more transmit antennas can increase the efficiency of energy beamforming, but incurs longer overhead for channel estimation. Thus, there exist an optimal number of transmit antennas, denoted by $m^{\star}$, which maximizes the the harvested energy, see Example~\ref{exa2}.
\end{myrem}%  (An example available in~\cite{YangHoGuan13}.)

\begin{myexa}\label{exa2}
Assuming full feedback, i.e., $G_{m,q}=2m$, we optimize $m$ according to Remark~\ref{remark1}. Define the set $\calK \triangleq \{ 0, 1, \cdots, T\}$. With $\tau=km$ for $k \in \calK$, from~\eqref{Energy_given_tau_L}, the harvested energy is $E(k,m)=\frac{m(T-km) (\sigma_z^2+k)}{m \sigma_z^2 + k}$, where $m=1,2,\cdots,\frac{T}{k}$. %, and $ m \in \mathbb{Z}^{+}$.%\lfloor \frac{T}{k} \rfloor
If $k=0$, we have $E(k,m)=T$, independent of $m$; thus there is no need to optimize $m$. Taking $m$ to be a continuous variable, given $k > 0$ and by taking the derivative of $E(k,m)$ with respect to $m$, the optimal number of transmit antennas $m^{\star}(k)$ is obtained as
\begin{align}
  m^{\star} (k) = \min \left\{ \frac{T}{k}, \frac{-k+\sqrt{k^2+ T \sigma_z^2}}{\sigma_z^2}\right\}.
\end{align}% \in \mathbb{Z}^{+}
%where the second term is positive.
Let $k_1= \arg \; \underset{k \in \calK }{\max} \; E \left(k, m^{\star} (k)\right)$. Due to the constraint that $m$ is an integer, the optimal number of transmit antennas is obtained by
\begin{align}
  m^{\star}&= \arg \max_{m \in \{\lfloor m^{\star} (k_1) \rfloor, \lceil m^{\star} (k_1) \rceil\} }   E \left(k_1, m\right).
\end{align}
\end{myexa}

\subsection{WPT with LMMSE Channel Estimator}\label{OptFixedLength_LMMSE}
Next, we consider the use of LMMSE, instead of LS, for the channel estimator.
\subsubsection{Correlated channel}
From~\eqref{eq:optbfvector}, the optimal beamforming vector is the largest eigenmode of $\bR_{\bh_q | \hatbh_q}$ in~\eqref{eq:CovMatrixCorrelatedLMMSE}. The total expected harvested energy is given in~\eqref{totalenergymiso2}. It is again difficult to analytically obtain the optimal preamble length. However, it can also be numerically obtained, which will be shown in Section~\ref{Section_NumericalResults}.

\subsubsection{Uncorrelated channel}
Using the optimal beamformer in~\eqref{eq:optbfvectorLMMSEUncorre}, the total harvested energy is obtained from~\eqref{totalenergymiso2} and~\eqref{LargestEigenvalueLMMSEUncorre}  as
\begin{align}
 E &= (T-\tau)  \left[  \frac{\sigma_e^2}{m + 2 \sigma_e^2} + \left( \frac{m +  \sigma_e^2}{m + 2 \sigma_e^2} \right)^2 \bbE_{\hatbh_q} \left( \left\| \hatbh_q \right\|_2^2\right) \right]. \label{totalenergymisoLMMSEUncorre}
\end{align}
An analogous result as Theorem~\ref{mythe1} can be obtained, since the total energy in~\eqref{totalenergymisoLMMSEUncorre} has a similar structure to~\eqref{totalenergymisoLSUncorre}. This will be numerically shown in Section~\ref{Section_NumericalResults}.

\section{Optimal Power Allocation} \label{SecOPA}
%Continuing from the derived optimal preamble length for the scenario of dynamic-length preamble and fixed-length preamble, respectively, in this section, we derive the optimal power allocation schemes for both scenarios.
Based on the derived optimal preamble length, in this section, we derive the optimal power allocation schemes for the scenario of dynamic-length preamble and fixed-length preamble, respectively.

\subsection{Dynamic-Length-Preamble Based Power Allocation}\label{DLPA}
By using dynamic-length preamble, the preamble length is typically shorter if the channel condition in one frame is good, and longer if the channel condition is bad. Intuitively, we can maximize the harvested energy by adjusting the transmit power for WPT, according to different channel conditions. In this section, we derive the optimal power allocation scheme, assuming the use of the optimal policy $\pi^{\star}$ for adapting the preamble length, although our subsequent results does not depend on the actual policy $\pi$ used.
%to Problem $\mathrm{P_1}$.

As in Section~\ref{Section-IV-A}, we take the channel estimate power in time slot $k$ as a random variable denoted by $V_k$, i.e., $V_k = \| \hatbh_k \|_2^2 \in \mathbb{R}^{+}$. Under policy $\pi^{\star}$, the preamble length, denoted by $\kappa$ time slots for convenience, is also random. When the receiver stops the channel estimation procedure at the end of the $\kappa$-th slot, we denote the corresponding channel estimate power by $\tilV_{\kappa}$, i.e., $\tilV_{\kappa} = \| \hatbh_{\kappa} \|_2^2 \in \calD_{\mathsf{s},\kappa}$.

\newcommand{\optV}{\tilV_{\kappa}}
\newcommand{\optv}{\tilv_{\kappa}}
\newcommand{\optp}{p(\optv, \kappa)}

First, we derive the joint pdf of $\kappa$ and $\optV$, denoted by $f(\optv, \kappa | \pi^{\star})$, upon using the optimal policy $\pi^{\star}$. For convenience, we omit the notation $\pi^{\star}$ in the sequel. We denote the joint pdf of $V_1,V_2,\cdots,V_{\kappa-1},\optV$ and $\kappa$ by $f(v_1,v_2,\cdots,v_{\kappa-1},\optv, \kappa)$. The joint pdf $f(\optv, \kappa)$ is given by the following recursive relation
\begin{align}
f(\tilv_1, 1) &=  \frac{\tilv_1^{m-1}}{2 \Gamma (m)(1+m \sigma_z^2)^{m-1}} \exp{\left(\frac{-\tilv_1}{1+m \sigma_z^2}\right)}, \label{eq:pdf_stopat1}
\end{align}%\nonumber \\
%&\qquad \qquad \qquad \; {\mathrm{for}} \; \tilv_1 \in \calD_{\mathsf{s},1} \kappa \!=\!
\begin{align}
  f(\optv, \kappa) &= \int_{v_{1} \in \calD_{\mathsf{c},1}} \cdots   \int_{v_{\kappa-1} \in \calD_{\mathsf{c},\kappa-1}}
  f(v_1,\cdots,v_{\kappa-1}, \optv, \kappa) d v_1 \; \cdots d v_{\kappa-1}, \label{eq:pdf_stopatk} \\
%  \end{align} %\quad {\mathrm{for}} \; \optv \in \calD_{\mathsf{s},\kappa}
  &\eqa \int_{v_1 \in \calD_{\mathsf{c},1}} \! f(v_1) \! \Bigg[\!  \int_{v_2 \in \calD_{\mathsf{c},2}}  f(v_2 | v_1) \bigg( \cdots \int_{{v_{\kappa-1} \in \calD_{\mathsf{c},\kappa-1}}} f(\optv | v_{\kappa-1}) f(v_{\kappa-1} | v_{\kappa-2}) d v_{\kappa_1} \! \cdots \bigg) d v_2 \Bigg] d v_1 \nonumber \\
%\begin{align}
  &\eqb \int_{v_1 \in \calD_{\mathsf{c},1}} \! f(v_1) \! \Bigg[\!  \int_{v_2 \in \calD_{\mathsf{c},2}}  f(v_2 | v_1) \bigg( \cdots \int_{{v_{\kappa-1} \in \calD_{\mathsf{c},\kappa-1}}} f(v_{\kappa} | v_{\kappa-1}) f(v_{\kappa-1} | v_{\kappa-2}) d v_{\kappa_1} \! \cdots \bigg) d v_2 \Bigg] d v_1, \nonumber
\end{align}%\substack
for $\kappa=2,\cdots,N-1$, where $\Gamma(\cdot)$ is the Gamma function. Here, (a) follows from the multiplication rule and Lemma~\ref{LemmaFirstOrderMarkov}, $f(v_1)$ is the same as~\eqref{eq:pdf_stopat1} with the argument replaced by $v_1 \in \bbR^{+}$, and $f(v_{i+1} | v_i) $ is given in~\eqref{eq:cond_pdf}, and (b) is from the fact $\optv=v_{\kappa}, \; \mathrm{for}\; v_{\kappa} \in \calD_{\mathsf{s},\kappa}$.
%the conditional pdf
%, and (b) is from the fact $\optv=v_{\kappa}, \; \mathrm{for}\; v_{\kappa} \in \calD_{\mathsf{s},\kappa}$.

\subsubsection{Optimal Length-and-Channel-Power Aware Power Allocation}\label{Sec2DPA}
In this section, we consider the scenario in which the power is allocated according to both the optimal preamble length $\kappa$ and the channel estimate power $\optv$. We refer to this scheme as length-and-channel-power aware power allocation (LCPA). With unit transmit power, the harvested energy is from~\eqref{eq:CondEnergy}%for convenience from~\eqref{eq:CondEnergy} Let the optimal preamble length be $\kappa$ time slots.
\begin{align}%\bE_{\text{cond}} \triangleq
 \tilE \left(\optv, \kappa \right)
  = m(N \!-\! \kappa) \left( \frac{m \sigma_z^2}{\kappa \!+\! m \sigma_z^2} \!+\! \frac{\kappa^2 \optv }{(\kappa \!+\! m \sigma_z^2)^2} \right).\label{eq:CondEnergy_optk}
\end{align}

We use $\optp$ to denote the transmit power for WPT in the frame with optimal preamble length $\kappa$ and channel estimate power $\optv \in \calD_{\mathsf{s},\kappa}$. We assume that  $\optp$ can be dynamically allocated, subject to the per-frame transmission power constraint $P_1$ and the average transmission power constraint $P_2$ over frames. To maximize the total expected harvested energy, we have the following optimization problem
%Given $k$ and $v_{k}$ (after feedback from the WP receiver), the transmitter allocate power $p_{k} (v_{k})$. % Correct objective function
\begin{subequations}
\label{eq:optimP4}
\begin{align}
   \mathrm{(J_1)} \ \ \underset{ \{\optp\}}{\text{max}} \ \ &\bbE_{\optv, \kappa} \left[ \tilE \left(\optv, \kappa \right) \optp \right] \label{eq:rewardJ1} \\
    \quad \text{s. t.} \ \
        &\bbE_{\optv, \kappa} \big[ m (N-\kappa) \optp \big] \leq P_2, \  \label{eq:const1J1} \\
        & 0 \leq \optp \leq P_1, \quad {\mbox{for}} \ \optv \! \in \! \calD_{\mathsf{s},{\kappa}},\ {\kappa}\!=\!0,\cdots, N\!-\!1. \label{eq:const2J2}
\end{align}
\end{subequations}
%\begin{subequations}
%\label{eq:optimP4}
%\begin{align}
%   \mathrm{(J_1)} \ \ \underset{ \{\optp\}}{\text{max}} \ \ &\bbE_{\optv, \kappa} \left[ \tilE \left(\optv, \kappa \right) \optp \right] \label{eq:rewardJ1} \\
%    \quad \text{s. t.} \ \
%        &\bbE_{\optv, \kappa} \big[ m (N-\kappa) \optp \big] \leq P_2, \  \label{eq:const1J1} \\
%        & 0 \leq \optp \leq P_1, \label{eq:const2J2}
%\end{align}
%\end{subequations}
%for $\optv \! \in \! \calD_{\mathsf{s},{\kappa}},\ {\kappa}\!=\!0,\cdots, N\!-\!1$.
In the constraint~\eqref{eq:const1J1}, the transmit power is utilized for WPT only in the last $(N-\kappa)$ slots.
%the constraint

Define $\eta(\optv,\kappa) = \frac{\tilE \left(\optv, \kappa \right)}{m(N-\kappa)}$. Here, $\eta(\optv,\kappa)$ is the efficiency of power transfer in the frame with optimal length $\kappa$ and channel estimate power $\tilv_{\kappa}$, which will be used as a criterion for adjusting the transmit power for WPT among frames. The optimal solution can be obtained by a greedy procedure as stated in Lemma~\ref{lem_2DPA}.
%As will be shown in the proof of Lemma~\ref{lem_2DPA}, the term $b(\optv,\kappa)$ is the ratio of the reward coefficient over the cost coefficient in a linear programming problem.

\begin{mylem}\label{lem_2DPA}
The optimal power allocation for Problem $\mathrm{J_1}$ is to allocate as much energy (up to $P_1$) to the frame with highest $\eta(\optv,\kappa)$ over all $\optv$ and all $\kappa$, then to the frame with the second highest $\eta(\optv,\kappa)$, and so on, until the average energy constraint $P_2$ is satisfied.
\end{mylem}

\begin{IEEEproof}
Define $a(\optv,\kappa) \!=\! m (N-{\kappa}) f(\optv, \kappa)$, and $x(\optv,\kappa) \!=\! a(\optv,\kappa) \optp$. Problem~$\mathrm{J_1}$ is rewritten as
\begin{subequations}\label{eq:optimP4a}
\begin{align}
   \underset{ \{x(\tilv_{\kappa}, \kappa)\}}{\text{max}} \ \
   & \sum_{\kappa=0}^{N-1} \left[\int_{\optv \in \calD_{\mathsf{s},{\kappa}}} \eta(\optv,\kappa) x(\optv,\kappa) d \optv \right] \\
    \quad \text{s. t.} \ \
        & \sum_{\kappa=0}^{N-1} \left[\int_{\optv \in \calD_{\mathsf{s},{\kappa}}} x(\optv,\kappa) d \optv \right] \leq P_2, \  \label{eq:SimplifiedJ1_P2}  \\
        & 0 \leq x(\optv,\kappa) \leq a (\optv, \kappa) P_1, \quad {\mbox{for}} \ \optv \! \in \! \calD_{\mathsf{s},{\kappa}},\ {\kappa}\!=\!0,\cdots, N\!-\!1.
\end{align}
\end{subequations}
%for $\optv \! \in \! \calD_{\mathsf{s},{\kappa}},\ {\kappa}\!=\!0,\cdots, N\!-\!1$.

We use $\bd_{\mathrm{s}}$ to denote the decreasing sorted vector of $\mathrm{vec} \left( \{\eta(\optv, \kappa)\} \right) $, where $\mathrm{vec} (\cdot)$ is the vectorization operator. Let $(\tilv_{\kappa_i}, \kappa_i)$ be the pair of the channel estimate power index and preamble length index associated to the $i$-th element in $\bd_{\mathrm{s}}$. The transformed problem is a linear programming problem. Its optimal solution is obtained as follows: For the first consecutive $(\tilde{\kappa}-1)$ slots, the power allocation is $P_1$; for slot $\tilde{\kappa}$, the the power allocation is such that the constraint~\eqref{eq:SimplifiedJ1_P2} is satisfied with equality; and for the remaining slots, no power is allocated. Here, $\tilde{\kappa}$ is chosen to be the maximally possible. This is because for $(\tilv_{\kappa_1}, \kappa_1)$, the objective function is increased the most, by setting the transmit power corresponding to $(\tilv_{\kappa_1}, \kappa_1)$ as the maximally possible, after which the transmit power for $(\tilv_{\kappa_2}, \kappa_2)$ is set as the maximally possible, and so on, until the average power constraint $P_2$ is satisfied.
\end{IEEEproof}

\subsubsection{Optimal Length-Aware Power Allocation}\label{SecLAPA}
To further reduce implementation complexity, we consider a simplified power allocation scheme in which the power is allocated according to only the optimal preamble length $\kappa$, referred as length-aware power allocation (LPA). Compared to the general Problem $\mathrm{J_1}$, we herein restrict $\optp=p(\kappa)$, independent of the channel estimate power $\optv$. As in the LCPA scheme, we employ the optimal policy $ \pi^{\star}$. Then, the probability that the optimal preamble length is $\kappa$, is given by
\begin{align}
  q_{\kappa} &=\int_{\optv \in \calD_{\mathsf{s},{\kappa}}} f(\optv, \kappa) d \optv.
\end{align}
With unit transmit power, the average harvested energy from frames of preamble length $\kappa$ is given by
\begin{align}
 \barQ_{{\mathrm{harv}}, \kappa} = \bbE_{\optv} \left[\tilE (\optv, \kappa) \right].
\end{align}
Problem $\mathrm{J_1}$ is thus simplified as
\begin{subequations}%\mathrm{(P_4)}
\label{eq:optimP5}
\begin{align}
   \mathrm{(J_{2})}\ \ \underset{ \{p({\kappa})\}}{\text{max}} \ \
   &\bbE_{\kappa} \left[p({\kappa}) \barQ_{{\mathrm{harv}}, \kappa} \right] \nonumber \\
     \quad \text{s. t.} \ \
        &\bbE_{\kappa} \left[m(N-\kappa) p({\kappa}) \right] \leq P_2, \nonumber \\
        & 0 \leq p({\kappa}) \leq P_1, \ {\mbox{for}} \ {\kappa}=0,\cdots, N-1. \nonumber
\end{align}
\end{subequations}
Let $\eta({\kappa}) = \frac{\barQ_{{\mathrm{harv}}, \kappa}}{m(N-\kappa)}$. Similar to Lemma~\ref{lem_2DPA}, the solution to Problem $\mathrm{J_{2}}$ is given without proof as in Lemma~\ref{cor_1DPA}.%$P_5$ Due to the similarity of

\begin{mylem}\label{cor_1DPA}
The optimal power allocation for Problem $\mathrm{J_2}$ is to allocate as much energy (up to $P_1$) to the frame with highest $\eta({\kappa})$ over all $\kappa$, then to the frame with the second highest $\eta(\kappa)$, and so on, until the average energy constraint $P_2$ is satisfied.
\end{mylem}

\subsection{Fixed-Length-Preamble Based Power Allocation}\label{PA_FixedLength_LS}%{OptFixedLength_LS}
In the fixed-length preamble scenario considered here, the optimal preamble length (i.e., $\kappa=\frac{\tau^{\star}}{m}$ time slots) is obtained in Section~\ref{WPT_MISO_Static}, and henceforth used for all frames. In this section, we consider the power allocation according to only the channel estimate power $\optv$, referred as channel-power-aware power allocation (CPA). For consistence, we use the same notations as the LCPA scheme in Section~\ref{Sec2DPA}, with difference here that the preamble length $\kappa$ is fixed.
%and~\ref{OptFixedLength_LMMSE}
%With unit transmit power, the harvested energy $\tilE_{\kappa}(\optv, \kappa)$ is obtained in~\eqref{eq:CondEnergy_optk}.

Let $p_{\kappa}(\optv)$ denote the transmit power for WPT in the frame with channel estimate power $\optv$. After obtaining $\optv$ via feedback, the WP transmitter performs WPT with transmit power $p_{\kappa}(\optv)$ in the current frame.
With the same power constraint $P_1$ and $P_2$ as in Problem $\mathrm{J_1}$, we formulate the following problem to maximize the harvested power
% Correct objective function
\begin{subequations}
\label{eq:optimP6}
\begin{align}
   \mathrm{(J_3)} \ \ \underset{ \{p_{\kappa}(\optv)\}}{\text{max}} \ \
   &\bbE_{\optv} \left[ \tilE_{\kappa}\left(\optv, \kappa)\right) p_{\kappa}(\optv) \right] \label{eq:rewardP6a} \\
    \quad \text{s. t.} \ \
        & m (N-\kappa) \bbE_{\optv} \left[ p_{\kappa}(\optv) \right] \leq P_2, \  \label{eq:const1P2a} \\
        & 0 \leq p_{\kappa}(\optv) \leq P_1, \ \mbox{for} \ \optv \in \bbR^{+}. \label{eq:const2P6a}
\end{align}
\end{subequations}
We note that given $\kappa$, the harvested energy $ \tilE_{\kappa}\left(\optv, \kappa\right)$ in~\eqref{eq:CondEnergy_optk} is a monotonically increasing function of the channel estimate power $\optv$. Similar to Lemma~\ref{lem_2DPA}, the solution to Problem $\mathrm{J_{3}}$ is given below without proof.

\begin{mylem}\label{cor_CPAPA}
The optimal power allocation for Problem $\mathrm{J_3}$ is to allocate as much energy (up to $P_1$) to the frame with highest channel estimate power $\optv$ over all $\optv$, then to the frame with the second highest $\optv$, and so on, until the average energy constraint $P_2$ is satisfied.
\end{mylem}%channel estimate power

\section{Numerical Results} \label{Section_NumericalResults}
In this section, we present numerical results to validate our results. We assume the time duration for the CE and WPT phases in each frame is $5 \textrm{ms}$, which consists of $T=126$ symbol periods. We set $m=3, N=42,\sigma_z^2 = -63 \textrm{dBm/Hz}$. We take the path loss model as $10^{-2} D^{-2}$, where the path loss exponent is 2, and $D = 10 \textrm{m}$ is the distance between the WP transmitter and WP receiver. A $20$dB path loss is assumed at a reference distance of $1 \textrm{m}$.

%\subsection{Harvested Energy Maximization}
%We herein give numerical results on wireless power transfer.

First, we simulate the harvested energy using the scheme based on the fixed-length preamble in Section~\ref{WPT_MISO_Static}, but without the adaptive power allocation in Section~\ref{PA_FixedLength_LS}. We fix the transmit power as $P_0=1$ watt.
%Section~\ref{OptFixedLength_LS} and Section~\ref{OptFixedLength_LMMSE}

\begin{figure} [t]
%\begin{subfigure}[b]{0.5\textwidth}
\centering
\includegraphics[width=.65\columnwidth] {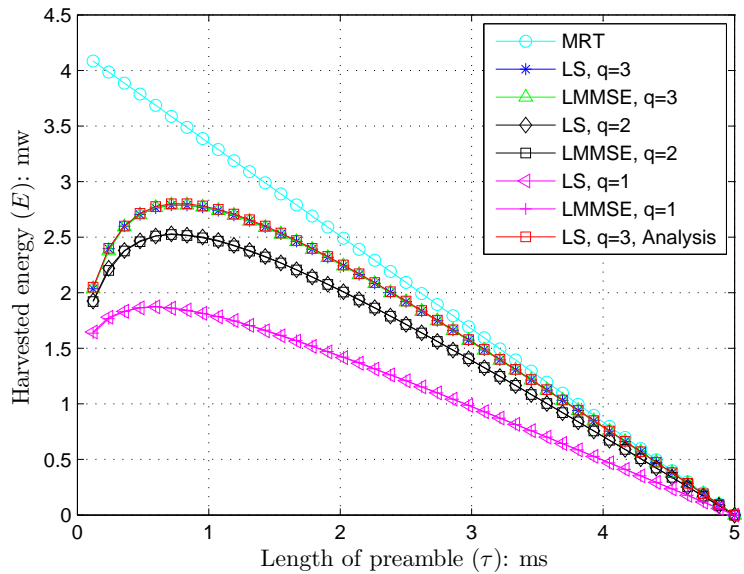}
%{figs/t0831b_varnoise1_uncorr_m3_20k.eps}
%\caption{Plot of MSE v.s. the length of preamble.}
\caption{Harvested energy in uncorrelated channel with MRT, LS and LMMSE.}
\label{fig:Fig8}%v.s. preamble length
%\end{subfigure}
%\begin{subfigure}[b]{0.5\textwidth}%
\end{figure}

We start from an uncorrelated MISO channel. Fig.~\ref{fig:Fig8} plots the harvested energy for different dimension $q$ of CSI feedback. With perfect CSI at the transmitter, the maximum ratio transmit (MRT) beamforming scheme harvests most energy, which provides an upper bound for all schemes that use fixed-length preamble. The $\square$-maker curve is plotted according to~\eqref{Energy_given_tau_L} in Theorem~\ref{mythe1} for different preamble length $\tau$. From~\eqref{eq:OPtLength} in Theorem~\ref{mythe1}, the optimal preamble length is $\tau^{\star}=0.72$ms, and the maximum harvested energy is $2.8$ milliwatts. The simulation results ($\ast$-maker curve) coincide with the analytical results. Moreover, the harvested energy is reduced as the dimension $q$ of CSI feedback decreases. Also, we observe that the LS based WPT achieves the same performance as the LMMSE-based WPT scheme as expected, since the channel is uncorrelated.

\begin{figure}
\centering
\includegraphics[width=.65\columnwidth] {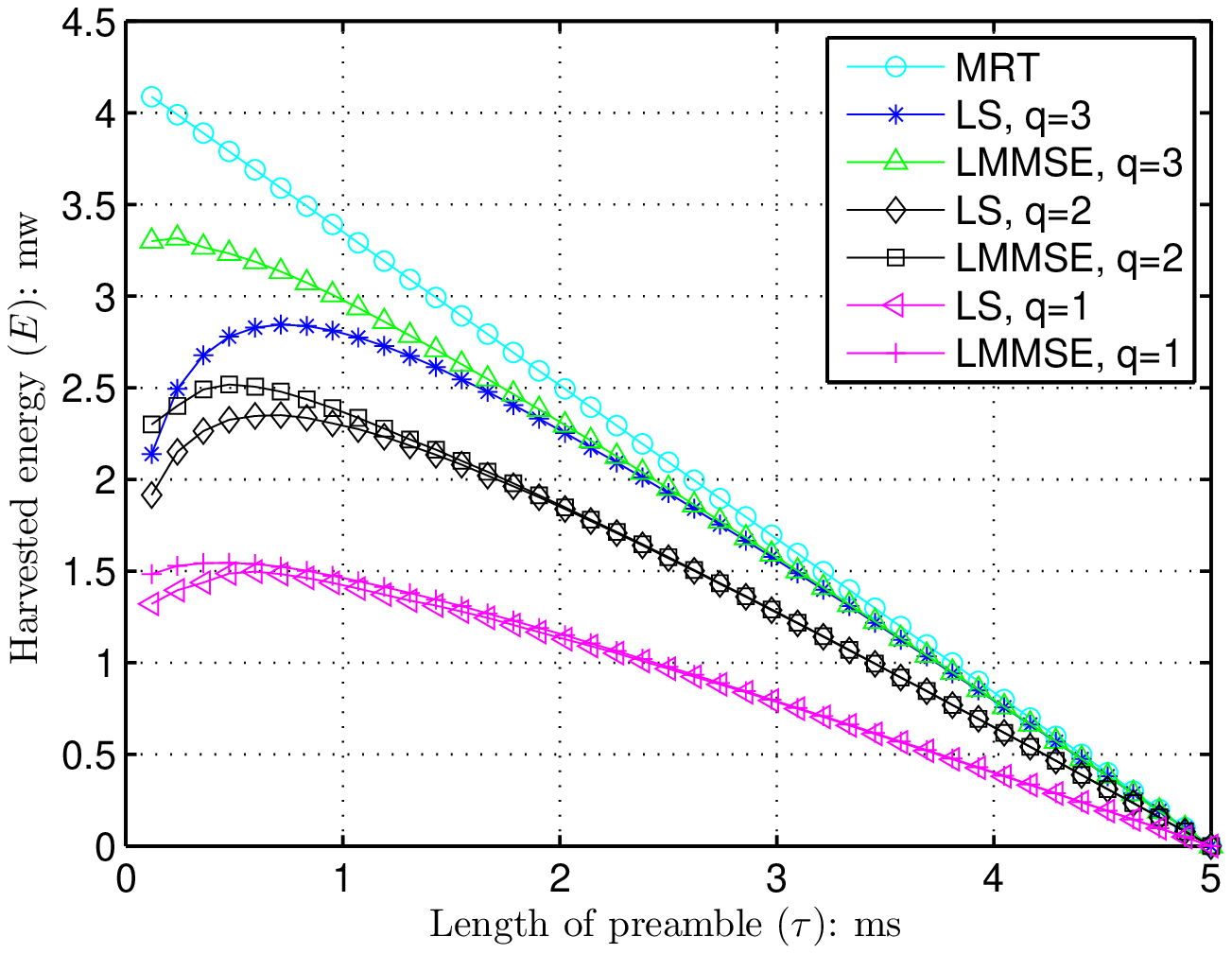}
%{figs/t0831a_varnoise1_corr_m3_20k.eps}
%\caption{Plot of MSE v.s. the length of preamble.}
\caption{Harvested energy in correlated channel with MRT, LS and LMMSE.}%v.s. preamble length
\label{fig:Fig9}
%\end{subfigure}%
%\caption{Plot of MSE v.s. the length of preamble.}
%\caption{Harvested energy v.s. length of preamble.}
\end{figure}

Next, we assume a correlated MISO channel, with channel correlation matrix that has the structure\footnote{This covariance matrix model is typical. See the literature~\cite{BigueshGershman06} and references therein.}: $[\bR]_{i,j}=\xi^{| i-j |}, \quad 0 \leq \xi <1$, where $i$ and $j$ are the indices of the entries. We set the correlation parameter $\xi=0.8$. The harvested energy is plotted in Fig.~\ref{fig:Fig9}. We observe that the LMMSE-based scheme transfers more energy than the LS-based WPT in general, due to the fact that an LMMSE estimator achieves more accurate channel estimation than an LS estimator.


%\begin{figure} [t]
%\begin{subfigure}[b]{0.5\textwidth}
%\centering
%\includegraphics[width=.99\columnwidth] {figs/t1014b_varnoise1_uncorr_m3_20k.eps}
%%{figs/t0831b_varnoise1_uncorr_m3_20k.eps}
%%\caption{Plot of MSE v.s. the length of preamble.}
%\caption{Harvested energy in uncorrelated channel.}
%\label{fig:Fig8}%v.s. preamble length
%\end{subfigure}
%\begin{subfigure}[b]{0.5\textwidth}%
%\centering
%\includegraphics[width=.99\columnwidth] {figs/t1014b_varnoise1_corr_m3_20k.eps}
%%{figs/t0831a_varnoise1_corr_m3_20k.eps}
%%\caption{Plot of MSE v.s. the length of preamble.}
%\caption{Harvested energy in correlated channel.}%v.s. preamble length
%\label{fig:Fig9}
%\end{subfigure}%
%%\caption{Plot of MSE v.s. the length of preamble.}
%\caption{Harvested energy v.s. length of preamble.}
%\end{figure}

\begin{figure} [t]
%\begin{subfigure}[b]{0.5\textwidth}
\centering
\hspace{-8mm}
\includegraphics[width=0.65\columnwidth] {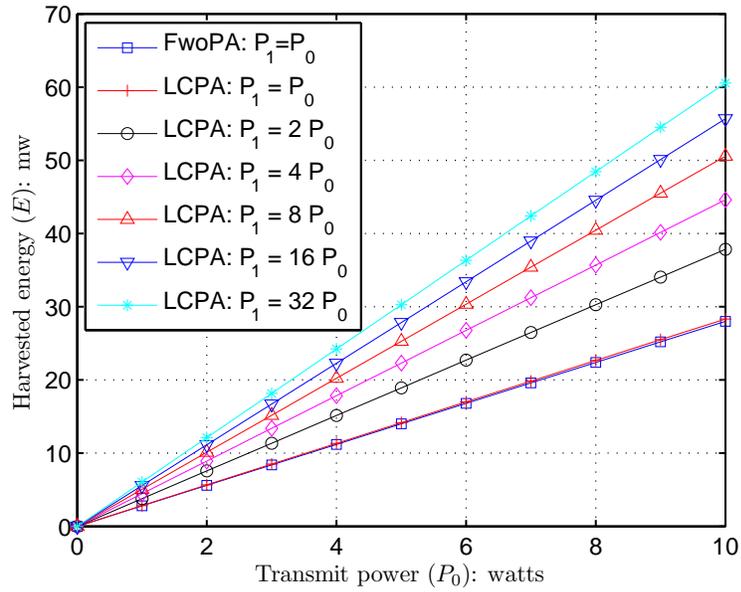}%t0927_E_vs_rangeA.eps
%\caption{Plot of MSE v.s. the length of preamble.}
\caption{Harvested energy of the LCPA and FwoPA scheme.}
\label{fig:Fig8c}
%\end{subfigure}
\end{figure}

Second, we compare the power-allocation based scheme with dynamic length preamble vs fixed-length preamble, under the same average energy consumption for WPT. Fig.~\ref{fig:Fig8c} compares the harvested energy by using dynamic-length preamble with the optimal length-and-channel-power aware power allocation (LCPA) in Section~\ref{Sec2DPA}, and that by using the the optimized fixed-length preamble without power allocation (FwoPA). In general, the harvested energy is proportional to the average transmit power $P_0$. Compared to the FwoPA scheme, the LCPA scheme achieves significant increase in harvested energy, especially when higher per-frame power $P_1$ is allowed. For instance, the harvested power is increased by $75\%$ for $P_1=8 P_0$. %, respectively. %by $56\%$  and P_1=4 P_0, \mbox{and}\

\begin{figure}
%\begin{subfigure}[b]{0.5\textwidth}%
\centering
\hspace{-8mm}
\includegraphics[width=0.65\columnwidth] {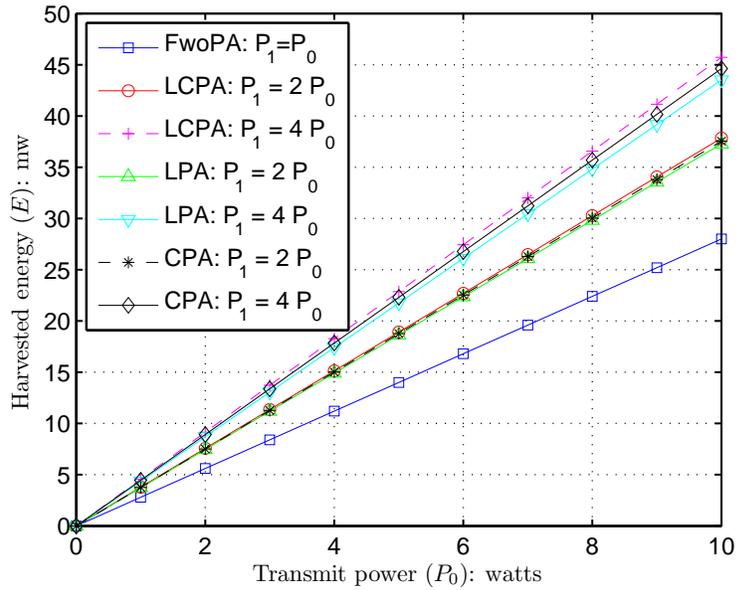}%t0927_E_vs_rangeA.eps
%\caption{Plot of MSE v.s. the length of preamble.}
%\caption{Comparison of different schemes.}
\caption{Comparison of the FwoPA, LCPA, LPA and CPA schemes.}
\label{fig:Fig8e}
%\end{subfigure}%
%\caption{Comparison of different schemes with/without power allocation.}
\end{figure}

Also, as shown in Fig.~\ref{fig:Fig8e}, we compare the LCPA scheme to the scheme based on dynamic-length preamble with length-aware power allocation (LPA) in Section~\ref{SecLAPA}, as well as the scheme based on the optimized fixed-length preamble with channel-power-aware power allocation (CPA) in Section~\ref{PA_FixedLength_LS}. It is observed that the CPA scheme and the LCPA scheme harvest almost the same amount of energy. This is because in the CPA scheme, the optimal preamble length is obtained after averaging all possible channel realizations, and the dynamical nature of the channels is fully exploited by the CPA scheme. Compared to the previous two schemes, the LPA scheme harvests less energy, since the dynamical nature of the channels is only partially exploited by the LPA scheme.


\section{Conclusion} \label{WPT_MISO_Conclusion}
The paper studies a MISO system where the transmitter delivers power to the receiver via energy beamforming, and the harvested energy is used to fulfill the need of the receiver to work. To maximize the harvested energy, we first derive the optimal energy beamformer. Then, we perform dynamic optimization for the preamble length, and also obtain the optimal preamble length offline to reduce the complexity. Moreover, we derive the optimal power allocation schemes for the wireless power transfer with dynamic-length preamble and fixed-length preamble, respectively. Future work extension will consider applications of the wireless power transfer system.

%Furthermore, we maximize the achievable net information rate with the harvested energy, by jointly optimizing the length of preamble and the time for uplink transmission.
%power the receiver for its uplink transmission

\appendices
%\section{Proof of Lemma~\ref{mylem1}}\label{AppCondPdf}
%\begin{IEEEproof}
%We note that both $\hatbh_q$ and $\bh_q$ are Gaussian vectors. Then using Bayes' Formula, the pdf of $\bh_q$ given $\hatbh_q$, is derived as
%
%\begin{align} \label{eq:condpdf_derive}
%  p(\bh_q | \hatbh_q) %&= \frac{p(\bh_q) p(\hatbh_q | \bh_q)}{p(\hatbh_q)} \nonumber \\
%  &= \frac{ \frac{1}{(2 \pi)^{m/2} | \bR_q |^{1/2}} \exp\left(-\frac{1}{2} \bh_q^H \bR_q^{-1} \bh_q \right) \cdot \frac{1}{(2 \pi)^{m/2} | \bR_{\be,q} |^{1/2}} \exp\left(-\frac{1}{2} (\bh_q - \hatbh_q)^H \bR_{\be,q}^{-1} (\bh_q - \hatbh_q) \right)} {\frac{1}{(2 \pi)^{m/2} | \bR_q + \bR_{\be,q} |^{1/2}} \exp\left(-\frac{1}{2} \hatbh_q^H (\bR_q+\bR_{\be,q} )^{-1} \hatbh_q \right) } \nonumber \\
%  &= {\frac{1}{(2 \pi)^{m/2} | (\bR_q^{-1}+\bR_{\be,q}^{-1})^{-1} |^{1/2}}} \cdot \nonumber \\
%  &\quad \exp\left(\!-\frac{1}{2} \left(\bh \!-\! ( \bR_{\be,q} \bR_q^{-1} \!+\! \bI_q )^{-1} \hatbh_q \right)^H (\bR_q^{-1}\! +\! \bR_{\be,q}^{-1}) \left(\bh_q \!-\! ( \bR_{\be,q} \bR_q^{-1} \!+\! \bI_q )^{-1} \hatbh_q \right) \right).
%\end{align}
%Clearly, \eqref{eq:condpdf_derive} is a Gaussian distribution with mean $( \bR_{\be,q} \bR_q^{-1} + \bI_q )^{-1} \hatbh_q$ and covariance $(\bR_q^{-1}+\bR_{\be,q}^{-1})^{-1}$.
%\end{IEEEproof}

%\newpage
\section{Proof for Lemma~\ref{LemmaFirstOrderMarkov}} \label{AppFirstOrderMarkov}
\begin{IEEEproof}
Let $\bh \!=\! [h_{1} \ h_{2} \ \cdots h_{m}]^T$, $\hatbh_r \!=\! [\hath_{r, 1} \ \hath_{r, 2} \ \cdots \hath_{r, m}]^T$ and $\hatbh_{r+1} \!=\! [\hath_{r+1, 1} \ \hath_{r+1, 2} \ \cdots \hath_{r+1, m}]^T$ be the channel vector, and two channel estimates obtained in time slots $r$ and $r+1$, respectively. From the property of LS channel estimation and the assumption that $\bh \sim \calC \calN\left(\mathbf{0}_m, \bI_m\right)$, we have that $\hatbh_r \sim \calC \calN\left(\mathbf{0}_m, \frac{r+m \sigma_z^2}{r} \bI_m\right)$, and $\hatbh_{r+1} \sim \calC \calN \left(\mathbf{0}_m, \frac{r+1+m \sigma_z^2}{r+1} \bI_m\right)$. Due to the assumption of independent elements over each vector, it suffices to consider the
pdf of channel coefficient $h$ (for some index) conditioned on two corresponding successive channel estimates $\hath_{r}$ and $\hath_{r+1}$ (for the same element index), denoted by $f(h | \hath_{r}, \hath_{r+1})$.

From~\eqref{eq:LSEstimate}, conditioned on $h$, the previous LS estimate is distributed as $\hath_r \sim \calC \calN (h, \frac{m \sigma_z^2}{r})$.
From~\eqref{eq:hkrecursive}, conditioned on $h$ and $\hath_r$, the current LS estimate is distributed as $\hath_{r+1} \sim \calC \calN \left(\frac{r}{r+1} \hath_r + \frac{h}{r+1}, \frac{m \sigma_z^2}{(r+1)^2}\right)$.
From Lemma~\ref{LemmaRecursiveCondDist}, conditioned on $\hath_r$, the current LS estimate is distributed as $\hath_{r+1} \sim \calC \calN \left(\frac{r(r+1+m \sigma_z^2)}{(r+1)(r+m\sigma_z^2)} \hath_r, \frac{m \sigma_z^2 (r+1+m \sigma_z^2)}{(r+1)^2(r+m\sigma_z^2)}\right)$.

The conditional distribution is thus
\begin{align}
f\left(h | \hath_{r}, \hath_{r+1}\right)&=\frac{f(h) f\left(\hath_r \left| h\right.\right) f\left(\hath_{r+1} \left| \hath_{r}, h\right. \right)}{f\left(\hath_{r}\right) f\left(\hath_{r+1} \left| \hath_{r}\right.\right)} \nonumber \\
%&= \frac{\frac{1}{\sqrt{2 \pi }} \exp \left( \frac{-h^2}{2} \right) \cdot \frac{1}{\sqrt{2 \pi \frac{m \sigma_z^2}{r}}} \exp \left[ \frac{-\left(\hath_{r}-h \right)^2}{2 \frac{m \sigma_z^2}{r}} \right] \cdot \frac{1}{\sqrt{2 \pi \frac{m \sigma_z^2}{(r+1)^2}}} \exp \left[ \frac{-\left(\hath_{r+1}-\frac{r}{r+1} \hath_r - \frac{h}{r+1}\right)^2}{2 \frac{m \sigma_z^2}{(r+1)^2}} \right]}{\frac{1}{\sqrt{2 \pi \frac{r+m \sigma_z^2}{r}}} \exp \left( \frac{-\hath_{r}^2}{2 \frac{r+m \sigma_z^2}{r}} \right) \cdot
%\frac{1}{\sqrt{2 \pi \frac{m \sigma_z^2 (r+1+m \sigma_z^2)}{(r+1)^2(r+m\sigma_z^2)}}} \exp \left[ \frac{-\left(\hath_{r+1}-\frac{r(r+1+m \sigma_z^2)}{(r+1)(r+m\sigma_z^2)} \hath_r\right)^2}{2 \frac{m \sigma_z^2 (r+1+m \sigma_z^2)}{(r+1)^2(r+m\sigma_z^2)}} \right]} \nonumber \\
&= \frac{1}{\sqrt{2 \pi \frac{m \sigma_z^2}{r+1+m \sigma_z^2}}} \exp \left[ \frac{-1}{2 \frac{m \sigma_z^2}{r+1+m \sigma_z^2}} \left(h-\frac{r+1}{r+1+m \sigma_z^2}\hath_{r+1}\right)^2 \right], \label{condPdfRelation2}
\end{align}
after some algebraic manipulations.
Clearly, the conditional distribution in~\eqref{condPdfRelation2} is Gaussian, which is independent of the previous channel estimate $\hath_r$.
% with mean $\frac{\hath_{r+1}}{1+ \sigma_{r+1}^2}$ and variance $\frac{\sigma_{r+1}^2}{1+ \sigma_{r+1}^2}$
%where (a) is from the Bayes' formula, and (b) is obtained
From Lemma~\ref{mylem1} for uncorrelated channel and $q=m$,
we then obtain that given $\hatbh_{r+1}$, the channel vector is distributed as $\bh \sim \calC \calN \left( \frac{\hatbh_{r+1}}{1+ \sigma_{r+1}^2} , \frac{\sigma_{r+1}^2}{1+ \sigma_{r+1}^2} \bI_m \right)$, where the error variance $\sigma_{r+1}^2=\frac{m\sigma_z^2}{r+1}$.
Hence, we have $f\left(h \left| \hath_{r}, \hath_{r+1} \right. \right)=f\left(h \left| \hath_{r+1}\right.\right)$.
Furthermore, we obtain by mathematical induction that
$f \left(h \left| h_1, h_2, \cdots, h_{k} \right. \right)=f\left(h \left|h_{k} \right. \right)$, for $k=1,\cdots, N-1$. The independence between elements completes this proof.
\end{IEEEproof}

\section{Proof for Lemma~\ref{LemOtpStopping}}\label{AppOptStopping}
\begin{IEEEproof} %\label{AppOptStopping}
We first consider the Policies~1 and 2, as follows.
Policy~1 has a decision sub-sequence $(\mathsf{c}, \mathsf{s}, \mathsf{c})$ over slots $r-1,r, r+1$. The corresponding states are $\bx_r, \bx_{r+1}$ and $\bx_{r+2}$, where $\bx_{r+1}=\bx_r$ because from \eqref{eq:state} the state value remains the same when $u_{r}=\mathsf{s}$.
Policy~2 is exactly the same policy as Policy~1, except that given state $\bx_r$ in slot $r$, Policy~2 performs CE followed by WP regardless of the state in slot $k+1$. Thus, the decision subsequence becomes $(\mathsf{c}, \mathsf{c}, \mathsf{s})$.
We aim to show that Policy~2 has strictly higher expected harvested energy than Policy~1. Both policies are statistically equivalent in slot $r+2$ and onwards because both have used the same number of slots for CE; hence the expected harvested energy in slot $r+2$ onwards are the same.
It thus suffices to compare the expected harvested energy of Policy~1 in slot $r$, denoted by $E_1^{r}(\bx_r)$, and that of Policy~2 in slot $r+1$, denoted by $E_2^{r+1}(\bx_r)$.
For the former case, the expected harvested energy is
%%$r,r+1, r+2$
%, i.e., the decision to select $u_{r+1}=\mathsf{s}$ at slot $k+1$ is based on the same state as used in slot $k$.
%that (i) performs CE for the first $r$ slots, then (ii) performs WP at slot $r+1$ based on state $\bx_r$, and (iii) performs CE at slot $r+2$.
\begin{align}
E_1^r \left(\hatbh_r\right) &= m \cdot \bE_{\bh |{\hatbh_r}} \left( \bw_{\mathrm{opt},r}^H \bh \bh^H \bw_{\mathrm{opt},r} \right)= m \left[\frac{m \sigma_z^2 }{(r+m \sigma_z^2)}+ \frac{r^2 \left\| \hatbh_r \right\|_2^2}{(r+m \sigma_z^2)^2} \right].
\label{eq:condenergy_stop_nextinstant}
\end{align}%{totalenergymiso2}
%where (a) comes from the fact $\hatbh_{r+a}$ has the same distribution as $\hatbh_{r+1}$ under policy $\pi_1$.
For the latter case, the channel estimate $\hatbh_{r+1}$ in the next slot $k+1$ is introduced. Hence the expectation for the harvested energy is taken over the conditional distribution $p(\bh,\hatbh_{r+1}|\hatbh_{r})=p(\hatbh_{r+1}|\hatbh_{r}) \cdot p(\bh|\hatbh_{r+1},\hatbh_{r}) $
\begin{align}
E_2^{r+1} \left(\bx_r\right) &\triangleq
m \cdot \bE_{\hatbh_{r+1} | {\hatbh_r}} \left[\bE_{\bh |{\hatbh_{r+1}}, \hatbh_r}\left( \bw_{\mathrm{opt},r+1}^H \bh \bh^H \bw_{\mathrm{opt},r+1} \right) \right]  \nonumber \\
&\eqa m \cdot \bE_{\hatbh_{r+1} | {\hatbh_r}} \left[\bE_{\bh |{\hatbh_{r+1}}}\left( \bw_{\mathrm{opt},r+1}^H \bh \bh^H \bw_{\mathrm{opt},r+1} \right) \right]  \nonumber \\
&\eqb m \left[\frac{m \sigma_z^2}{r+1 + m \sigma_z^2} + \frac{(r+1)^2 \bE_{\hatbh_{r+1} | {\hatbh_r}}  \left\| \hatbh_{r+1} \right\|_2^2}{(r+1+m \sigma_z^2)^2} \right] \nonumber \\
&\eqc m \left[\frac{m \sigma_z^2 (r+m+m \sigma_z^2)}{(r+m \sigma_z^2) (r+1+m \sigma_z^2)} + \frac{r^2 \left\| \hatbh_r \right\|_2^2}{(r+m \sigma_z^2)^2} \right],  \label{eq:condenergy_policy1}
\end{align}
where (a) comes from Lemma~\ref{LemmaFirstOrderMarkov}, (b) follows \eqref{eq:largesteigenvalue}, and (c) is from the conditional mean in~\eqref{eq:Mean_NoncentralChiSquare}.
%\eqref{eq:cond_pdf}.
We conclude that Policy~1 is strictly worse than Policy~2, since $E_2^{r+1}(\bx_r)-E_1^{r}(\bx_r)=\frac{m^2(m-1) \sigma_z^2}{(r+m \sigma_z^2)(r+1+m\sigma_z^2)}>0$.

The same argument extends to the case if the decision subsequence of Policy~1 is of the structure $(\mathsf{c}, \mathsf{s}, \cdots, \mathsf{s}, \mathsf{c})$, i.e., there are more than one slot with decision $\mathsf{s}$ in between the two slots with decision $\mathsf{c}$.
Moreover, the same argument holds if Policy~1 is of the structure $(\mathsf{s}, \cdots, \mathsf{s}, \mathsf{c})$, by treating the case without CE as a special case with CE but with estimation error. Lemma~\ref{LemOtpStopping} must then hold; otherwise, there exists a decision subsequence with a structure that was shown to be suboptimal.
%
%Clearly, it holds that $E_1 (\hatbh_r)> E_2 (\hatbh_r)$. Because under police $\pi_2$, the energy beamforming from slot $r+1$ to slot $r+a-1$ is based on $\hatbh_{r}$, which is less accurate than the channel estimate $\hatbh_{r+1}$ under policy $\pi_2$. By reasoning, we conclude that among all polices whose realizations has the same total length of preamble $k^*$, the realization of the optimal policy will always perform CE in the first $k^*$ time slots. Hence, the optimal policy has the structure claimed in this lemma.
\end{IEEEproof}

\section{Proof of Theorem~\ref{mythe1}}\label{AppendixB}
\begin{IEEEproof}
For the $q$-dimensional feedback, the receiver feeds back $\hatbh_q$ and the corresponding indices set $\calI$ to the transmitter. From~\eqref{totalenergymiso2} and~\eqref{eq:largesteigenvalue}, using the optimal beamformer in~\eqref{eq:optbfvector}, the total harvested energy is %of cardinality $q$,
\begin{align}
\hatE(\beta) &= (T- \beta m^2) \left( \frac{\sigma_z^2}{\beta+\sigma_z^2} + \frac{\beta^2 \sum \nolimits_{i=1}^q  \bbE_{\hath_{(i)}} \left( \left| \hath_{(i)} \right|^2 \right) }{(\beta+\sigma_z^2)^2} \right). \nonumber %\label{eq:partialFBEnergy}
\end{align}

Assuming independent Rayleigh fading channels, we have that the channel estimates $\hath_i$'s are independent zero-mean complex Gaussian random variables with variance $(1 +\frac{\sigma_z^2}{\beta})$. Denote $\bu=\frac{2}{1+{\sigma_z^2}/{\beta}} \left[| \hath_1|^2 \;  \dots  \; | \hath_m|^2\right]^T$. Elements of the random vector $\bu$ are thus independent Chi-Square random variables. %| \hath_2|^2 \;

Let $u_{(r)}$ denote the random variable corresponding to the $r$-th largest observation of the $m$ original random variables. Actually, $u_{(r)}$ is the $r$-th order statistics. From order statistics, the pdf of $u_{(r)}$ is given by%{\emph{}}%, i.e., $u_{(1)} \geq u_{(2)} \geq \cdots \geq u_{(m)}$
\begin{align}
  p_{u_{(r)}}=\frac{m!}{2 (m-r)! (r-1)!} e^{-\frac{r u}{2}} \left(1-e^{-\frac{u}{2}}\right)^{m-r}. \nonumber
\end{align}
Denote $C_{m,r}=\frac{m!}{(m-r+1)! (r-1)!}$. The expectation of $u_{(r)}$ is further derived as
\begin{align}
  \bbE \left(u_{(r)}\right) &= \int_0^{\infty} u \frac{m!}{2 (m-r)! (r-1)!} e^{-\frac{r u}{2}} \left(1-e^{-\frac{u}{2}}\right)^{m-r} du \nonumber \\
  %&=\frac{m!}{(m-r+1)! (r-1)!} \int_0^{\infty} u e^{- \frac{(r-1)u}{2}} d \left(1-e^{\frac{u}{2}} \right)^{m-r+1} \nonumber \\
  &= C_{m,r}\left[ u e^{- \frac{(r-1)u}{2}} \left(1-e^{\frac{u}{2}} \right)^{m-r+1} - \int \left(1-e^{\frac{u}{2}}\right)^{m-r+1} d \left( u e^{- \frac{(r-1)u}{2}} \right)\right] \bigg|_0^{\infty} \nonumber \\
  %&= \frac{m!}{(m-r+1)! (r-1)!}  \bigg[ u e^{- \frac{(r-1)u}{2}} \left(1-e^{\frac{u}{2}} \right)^{m-r+1} \nonumber \\
  %&\quad - \int e^{- \frac{(r-1)u}{2}} \left(1-e^{\frac{u}{2}}\right)^{m-r+1} du - \frac{1-r}{2} \int  u e^{- \frac{(r-1)u}{2}} \left(1-e^{\frac{u}{2}} \right)^{m-r+1} du \bigg] \bigg|_0^{\infty} \nonumber \\
  &= C_{m,r} \bigg[ u e^{- \frac{(r-1)u}{2}} \left(1-e^{\frac{u}{2}} \right)^{m-r+1} - \int e^{- \frac{(r-1)u}{2}} \left( 1+\sum \nolimits_{s=1}^{m-r+1} { {m-r+1} \choose s} (-1)^s e^{-\frac{s}{2}u} \right)du \nonumber \\
  &\quad - \frac{1-r}{2} \int  u e^{- \frac{(r-1)u}{2}} \left( 1+\sum \nolimits_{s=1}^{m-r+1} { {m-r+1} \choose s} (-1)^s e^{-\frac{s}{2}u} \right) du \bigg] \bigg|_0^{\infty} \nonumber \\
  &=C_{m,r}  \bigg[\! \sum \nolimits_{s=1}^{m \!-\!r\!+\!1} { {m\!-\!r\!+\!1} \choose s} (-1)^s e^{-\frac{s+r-1}{2}u} \big( u \!+\! \frac{2}{r\!+\!s\!-\!1} \!-\! \frac{r-1}{r\!+\!s\!-\!1}u \!-\! \frac{2 (r-1)}{(r\!+\!s\!-\!1)^2} \big) \bigg]\bigg|_0^{\infty} \nonumber \\
  %&= \frac{2 m!}{(m-r+1)! (r-1)!} \sum \nolimits_{s=1}^{m-r+1} { {m-r+1} \choose s} \frac{s (-1)^{s+1}}{(s+r-1)^2} \nonumber \\
  &= \frac{2 m!}{(r-1)!} \sum \nolimits_{s=1}^{m-r+1} \frac{s (-1)^{s+1}}{(m-r+1-s)! s!  (r+s-1)^2}. \nonumber
\end{align}

Denote $G_{m,q} \triangleq  \sum \nolimits_{r=1}^q \bbE \left(u_{(r)}\right)$, which is given in the Table of Appendix~\ref{AppendixC}. We have that $g_{m,1}$ is no less than than 2. Moreover, $G_{m,m}=2m$, since $\sum \nolimits_{r=1}^m \bbE \left(u_{(r)}\right)$ is the variance of a ($m$ degrees of freedom) Chi-Square random variable. Then we obtain the total harvested energy %Table\ref{table1} in g_{m,r}$, where $g_{m,r} = , for all $m=1,2,\cdots$
\begin{align}
\hatE \left(\beta\right) = (T-\beta m^2)  \frac{G_{m,q} \beta + 2 \sigma_z^2}{2 \left(\beta+\sigma_z^2\right)}.
\end{align}

Moreover, the first-order derivative and the second-order derivative of $\hatE \left(\beta\right) $ yield as
\begin{align}
  \hatE' \left(\beta\right)  &=- \frac{m^2 G_{m,q}}{2} \frac{(\beta+\sigma_z^2)^2 + \frac{\sigma_z^2}{m^2 G_{m,q}} (m^2 \sigma_z^2 + T)(2-G_{m,q}) }{(\beta+\sigma_z^2)^2}, \\
    \hatE'' \left(\beta\right)  &=- \frac{\sigma_z^2 (m^2 \sigma_z^2 +T) (G_{m,q}-2)}{(\beta+\sigma_z^2)^3}.
\end{align}
Let $\beta_1$ and $\beta_2$ be the roots of $\hatE'(\beta)=0$. We have
\begin{align}
  \beta_{1,2} &= - \sigma_z^2 \pm \sqrt{ \frac{\sigma_z^2}{m^2 G_{m,q}} (m^2 \sigma_z^2 + T)(G_{m,q}-2)}
\end{align}

Since $\beta_2$ is always negative, it is useless for the analysis. When $\sigma_z^2 \leq \frac{T(G_{m,q} - 2)}{2 m^2}$, we have $\beta_1$ is always positive, and $\hatE' \left(\beta\right) > 0 \ (<0), \ \forall \ 0 \leq \beta \leq \beta_1 \ \left(\beta_1 \leq \beta \leq T/m^2\right)$. Moreover, $\hatE  \left(\beta\right) $ is a concave function for $\beta \geq 0$, since $\hatE''  \left(\beta\right)  <0, \ \forall \ \beta \geq 0$. Hence,  $\beta_1$ is the unique value of $\beta$ that maximizes $\hatE \left(\beta\right)$. Clearly, $\hatE \left(\beta\right)$ is maximized at $\beta=0$, when $\sigma_z^2 > \frac{T(G_{m,q} - 2)}{2 m^2}$, or equivalently, $\beta_1 < 0$. For convenience, define the function $E(\tau) \triangleq \hatE'\left(\tau/m^2\right) $, and set $\tau_1=m^2 \beta_1$.

Due to the constraint that the preamble length should be multiples of the number of transmit antennas $m$, we obtain the optimal preamble length as $\tau^{\star}= \arg \underset{\tau \in \{\lfloor \tau_1 \rfloor, \lceil \tau_1 \rceil\} }{\max} E \left(\tau \right)$, if $\sigma_z^2 \leq \frac{T(G_{m,q} - 2)}{2 m^2}$; and $\tau^{\star}=0$, if $\sigma_z^2 > \frac{T(G_{m,q} - 2)}{2 m^2}$.
Thus, the maximum harvested energy $E_{\max}=E \left(\tau^{\star}\right)$.
\end{IEEEproof}

\section{Table for $G_{m,q}$} \label{AppendixC}
\begin{table}[htbp] %\label{table1}%
%\caption{Values of $G_{m,q}$}
\centering
\small{
\begin{tabular}{c c c c c c c c c c c}
  \hline \hline
   $G_{m,q}$ & q=1 & 2 & 3  & 4  & 5  & 6  &  7 &  8 & 9  & 10   \\
  \hline
  m=1 & 2 & \  & \  & \  & \  & \  & \  & \  & \  & \  \\
  2 & 3 & 4 & \ & \ & \ & \ & \ & \ & \ & \ \\
  3 & 3.6667 & 5.3333 & 6 & \ & \ & \ & \ & \ & \ & \ \\
  4 & 4.1667 & 6.3333 & 7.5000 & 8 & \ & \ & \ & \ & \ & \ \\
  5 & 4.5667 & 7.1333 & 8.7000 & 9.6000 & 10 & \ & \ & \ & \ & \ \\
  6 & 4.9000 & 7.8000 & 9.7000 & 10.9333 & 11.6667 & 12 & \ & \ & \ & \ \\
  7 & 5.1857 & 8.3714 & 10.5571 & 12.0762 & 13.0952 & 13.7143 & 14 & \ & \ & \ \\
  8 & 5.4357 & 8.8714 & 11.3071 & 13.0762 & 14.3452 & 15.2143 & 15.7500 & 16 & \ & \ \\
  9 & 5.6579 & 9.3159 & 11.9738 & 13.9651 & 15.4563 & 16.5476 & 17.3056 & 17.7778 & 18 & \ \\
  10 & 5.8579 & 9.7159 & 12.5738 & 14.7651 & 16.4563 & 17.7476 & 18.7056 & 19.3778 & 19.8000 & 20 \\
  \hline
\end{tabular}}
\end{table}

\renewcommand{\baselinestretch}{1.45}
%% Reference
\bibliography{IEEEabrv,reference1308}
\bibliographystyle{ieeetr}

\end{document}